\documentclass[journal,draftcls,onecolumn,12pt,twoside]{IEEEtranTCOM}
\usepackage{amsmath,amsfonts}
\usepackage{amsmath}
\usepackage{gensymb}
\usepackage{algorithmic}
\usepackage{array}
\usepackage{textcomp}
\usepackage{stfloats}
\usepackage{url}
\usepackage{verbatim}
\usepackage{graphicx}
\usepackage{physics}
\usepackage{subcaption}
\usepackage{booktabs}
\newcommand{\opan}[1]{\mathop{}\!{\hat{#1}}}
\newcommand{\opdag}[1]{\mathop{}\!{\hat{\rule{0pt}{1.5ex}#1}^{\dag}}}
\def\BibTeX{{\rm B\kern-.05em{\sc i\kern-.025em b}\kern-.08em
    T\kern-.1667em\lower.7ex\hbox{E}\kern-.125emX}}
\usepackage{balance}
\begin{document}
\bstctlcite{IEEEexample:BSTcontrol}
\title{Satellite-to-Ground Continuous Variable Quantum Key Distribution: The Gaussian and Discrete Modulated Protocols in Low Earth Orbit}
\author{M. Sayat, B. Shajilal, S. P. Kish, S. M. Assad, T. Symul, P. K. Lam, N. Rattenbury, J. Cater}
\maketitle
\vspace{-25mm}
\begin{abstract}
The Gaussian modulated continuous variable quantum key distribution (GM-CVQKD) protocol is known to maximise the mutual information between two parties during quantum key distribution (QKD). An alternative modulation scheme is the discrete modulated CVQKD (DM-CVQKD) protocol. In this paper, we study the Phase Shift Keying ($M$-PSK) and Quadrature Amplitude Modulation ($M$-QAM) DM-CVQKD protocols along with the GM-CVQKD protocol over a satellite-to-ground link in the low SNR regime. We use a satellite-to-ground link model which takes into account geometric losses, scintillation, and scattering losses from the link distance, atmospheric turbulence, and atmospheric aerosols, respectively. In addition, recent multidimensional (MD) and multilevel coding and multistage decoding (MLC-MSD) reconciliation method models in combination with multiedge-type low-density parity-check (MET-LDPC) code models have been used to determine the reconciliation efficiency. The results show that GM-CVQKD outperforms DM-CVQKD. In addition, GM-CVQKD with MD reconciliation outperforms GM-CVQKD with MLC-MSD reconciliation in the finite size limit by producing positive secret key rates at larger link distances and lower elevation angles.
\end{abstract}
\vspace{-5mm}
\section{Introduction}
\IEEEPARstart{Q}{uantum} key distribution (QKD)~\cite{Lo1999} is a method of sharing a secret key between two parties, Alice and Bob, where eavesdropping by Eve can be inferred as a consequence of fundamental quantum mechanics. The most significant advancement of QKD in space-based applications was the demonstration of discrete variable QKD (DVQKD), which uses the available degrees of freedom of single photons, to encode a key between optical ground stations on Earth and the Micius satellite~\cite{Yin2020}. Despite its successful deployment in space, its use of expensive and inefficient single photon detectors for detection poses challenges for its popularisation and commercialisation~\cite{Laudenbach2018}. An appealing alternative is continuous variable QKD (CVQKD) which uses multi-photon technologies to encode the key in the continuous X and P quadratures of light~\cite{Laudenbach2018, Garcia2006}. Its use of homodyne or heterodyne detection is more cost-effective, more compatible with standard telecommunication optical networks, and more efficient, offering higher secret key rates. 

CVQKD experiments have predominantly been restricted to fibre-based systems in the laboratory where a secret key rate of 14.2~Mbit/s over 15~km of optical fibre has been demonstrated using a local local oscillator~\cite{Ren2020}. The first demonstration of fibre-based CVQKD over 100 km was performed by controlling and suppressing excess noise \cite{Huang2016}. A secret key rate of 2.1~Gbit/s in free space has been achieved using 10-channel wavelength division multiplexing in the weak turbulence regime~\cite{Qu2018}, and a secret key rate greater than 1.68~Gbit/s can be reached in a lossless and excess noise free system that uses two polarisations, six wavelengths, and four orbital angular momentum for multiplexing~\cite{Qu2017}. A study of atmospheric effects on quantum communications over 1.6~km of free-space was performed and the experimental setup was capable of CVQKD~\cite{Heim2014}. However, there has only been one demonstration of free-space CVQKD which occurred over 460~m in an urban environment and achieved a secret key rate of 0.152~kbit/s using polarised coherent states with uni-dimensional Gaussian modulation~\cite{Shen2019}.

The two modulation approaches for CVQKD to encode a key are Gaussian modulation (GM-CVQKD)~\cite{Laudenbach2018} and discrete modulation (DM-CVQKD)~\cite{Leverrier2009}, where implementations can be found in associated publications, \cite{Grosshans2002, Shen2010}. The GM-CVQKD protocol is capable of achieving higher secret key rates as the coherent state distribution follows a continuous Gaussian distribution allowing for more positions on the optical phase space \cite{Laudenbach2018}. In contrast, DM-CVQKD positions a finite number of coherent states on the optical phase space, limiting its achievable SKR \cite{Zhao2020, Zhang2012, Becir2010}. As DM-CVQKD coherent state distribution does not follow a continuous distribution, but a discretised one, it can be argued that it can accommodate more excess noise.

In this work, the feasibility of low Earth orbit (LEO - orbit altitudes of 160-1000~km) satellite-to-ground GM-CVQKD and DM-CVQKD is investigated in the asymptotic and finite size limits. The DM-CVQKD protocols studied are the $M$-PSK protocol that assumes Gaussian optimality under collective attacks~\cite{Leverrier2009, Zhao2020, Zhang2012, Becir2010, Djordjevic2019, Wang2019}, and the $M$-QAM protocol which does not assume Gaussian optimality under collective attacks \cite{Denys2021}. The work is structured as follows. Section II introduces the GM-CVQKD protocol as well as the $M$-PSK and $M$-QAM DM-CVQKD protocol. Section III describes the process used when considering finite size effects. Section IV introduces the LEO satellite to optical ground station (OGS) link model. Section V discusses the achievable secret key rates of LEO satellite-to-ground GM-CVQKD and DM-CVQKD, and the resulting trends in SKR from parameter variation. Section VI presents the conclusions.

\section{The GM-CVQKD and DM-CVQKD protocols}
In this section, the GM-CVQKD protocol and the $M$-PSK DM-CVQKD protocol with security under collective attacks and assuming Gaussian optimality are investigated. The $M$-QAM DM-CVQKD protocol with security under collective attacks without assuming Gaussian optimality is also studied.

\subsection{GM-CVQKD}
In GM-CVQKD \cite{Laudenbach2018}, Alice sends a series of displaced coherent states,

\begin{equation}
    \ket{\alpha} = \ket{q + ip}, 
\end{equation}

where $q$ and $p$ are the amplitude and phase quadrature components. These are both random variables of the same zero-centred normal distribution:

\begin{equation}
    q,p = \mathcal{N}(0, V_A).
\end{equation}

Here, $V_A$ is the modulation variance. The covariance matrix describing the Gaussian modulated coherent state sent from Alice to Bob is

\begin{equation}
    \label{covariancematrix}
    \gamma_{AB} = 
    \begin{bmatrix} (V_A + 1)\boldsymbol{\mathrm{I}} & \sqrt{T(V_A^2 + 2V_A)}\boldsymbol{\sigma_z} \\ \sqrt{T(V_A^2 + 2V_A)}\boldsymbol{\sigma_z} &  T(V_A + 1 + \chi_{\mathrm{line}})\boldsymbol{\mathrm{I}} \end{bmatrix},
\end{equation}
where $V_A$ is the modulation variance of Alice, $T$ is the overall transmittance between Alice and Bob, and $\chi_{\textrm{line}}$ is the noise in the channel line expressed in shot noise units. \textbf{I} and $\boldsymbol{\mathrm{\sigma_z}}$ are the identity matrix, $\begin{bmatrix} 1 & 0 \\ 0 & 1 \end{bmatrix}$, and the Pauli matrix, $\begin{bmatrix} 1 & 0 \\ 0 & -1 \end{bmatrix}$, respectively.

The asymptotic limit secret key rate (SKR) [bits/pulse] is calculated as
\begin{equation}
\label{SKR}
\mathrm{SKR}_{\mathrm{asy}} = \beta I_{AB} - S_{BE},
\end{equation}
where $\beta$ is the reconciliation efficiency, $I_{AB}$ is the mutual information between Alice and Bob, and $S_{BE}$ upper bounds the Holevo information that represents the maximum mutual information between Eve and Bob in the protocol~\cite{Leverrier2009}. The mutual information for homodyne and heterodyne detection are calculated as 
\begin{equation}
    \label{MutualInformation}
    \begin{aligned}
        I_{AB,\mathrm{hom}} &= \frac{1}{2}\log_2 \frac{(V_A + 1) + \chi_{\mathrm{tot}}}{1 + \chi_{\mathrm{tot}}}, \\
        I_{AB,\mathrm{het}} &= \log_2 \frac{(V_A + 1) + \chi_{\mathrm{tot}}}{1 + \chi_{\mathrm{tot}}},
    \end{aligned}
\end{equation}
where the total excess noise, $\chi_{\mathrm{tot}}$, combines both the channel noise, $\chi_{\mathrm{line}} = \frac{1}{T} - 1 + \epsilon_{\mathrm{ch}}$ (where $\epsilon_{\mathrm{ch}}$ is the channel excess noise (Table \ref{ExcessNoise})), and detection excess noise, $\chi_{\mathrm{hom/het}}$, and is expressed as $\chi_{\mathrm{tot}} = \chi_{\mathrm{line}} + \frac{\chi_{\mathrm{hom/het}}}{T}$~\cite{Zhang2012}. The detection noise is different for homodyne and heterodyne detection where $\chi_{\mathrm{hom}} = \frac{(1 - \eta) + \epsilon_{\mathrm{det}}}{\eta}$ and $\chi_{\mathrm{het}} = \frac{1 + (1 - \eta) + 2\epsilon_{\mathrm{det}}}{\eta}$, respectively. Here, $\eta$ is the detector efficiency and $\epsilon_{\mathrm{det}}$ is the detector excess noise (Table \ref{ExcessNoise}). 

The Holevo information is calculated as
\begin{equation}
\label{HolevoBound}
    \begin{aligned}
        S_{BE} = G\left(\frac{\lambda_1 - 1}{2}\right) + G\left(\frac{\lambda_2 - 1}{2}\right)
        - G\left(\frac{\lambda_3 - 1}{2}\right) - G\left(\frac{\lambda_4 - 1}{2}\right),
    \end{aligned}
\end{equation}
where $G(x) =  (x+1)\log_2 (x+1) -x\log_2 x$ and $\lambda$ are the symplectic eigenvalues of the covariance matrix, $\gamma_{AB}$. $\lambda_{1,2}$ is calculated as
\begin{equation}
    \begin{aligned}
        \lambda_{1,2} = \sqrt{\frac{1}{2} (A \pm \sqrt{A^2 - 4B})},
\end{aligned}
\end{equation}
where
\begin{equation}
    \begin{aligned}
        &A = (V_A + 1)^2 +T^2(V_A + 1 + \chi_{\mathrm{line}})^2 - 2TZ^2, \;\mathrm{and} \\
        &B = (T(V_A + 1)^2 + T(V_A + 1)\chi_{\mathrm{line}} - TZ^2)^2.
\end{aligned}
\end{equation}
$\lambda_{3,4}$ is calculated as
\begin{equation}
    \begin{aligned}
        \lambda_{3,4} = \sqrt{\frac{1}{2} (C \pm \sqrt{C^2 - 4D})}, \\
    \end{aligned}
\end{equation}
where
\begin{equation}
\begin{aligned}
    &C_{\mathrm{hom}} = \frac{A\chi_{\mathrm{hom}} + (V_A + 1)\sqrt{B} + T(V_A + 1 + \chi_{\mathrm{line}})}{T(V_A + 1 + \chi_{\mathrm{tot}})}, \\
    &D_{\mathrm{hom}} = \sqrt{B} \frac{V_A + 1 + \sqrt{B}\chi_{\mathrm{hom}}}{T(V_A + 1 + \chi_{\mathrm{tot}})}, \\
\end{aligned}
\end{equation}
for homodyne detection, and
\begin{equation}
\label{Chet_Dhet}
\begin{aligned}
    &C_{\mathrm{het}} = \frac{A\chi_{\mathrm{het}}^2 + B + 1 + 2TZ^2}{[T(V_A + 1 + \chi_{\mathrm{tot}})]^2}, \\
    &+ \frac{2\chi_{\mathrm{het}}[(V_A + 1)\sqrt{B} + T(V_A + 1 + \chi_{\mathrm{line}})]}{[T(V_A + 1 + \chi_{\mathrm{tot}})]^2}, \\
    &D_{\mathrm{het}} = \left(\frac{V_A + 1 + \sqrt{B}\chi_{\mathrm{het}}}{T(V_A + 1 + \chi_{\mathrm{tot}})}\right)^2
\end{aligned}
\end{equation}
for heterodyne detection. Here, $Z = \sqrt{V_A^2 + 2V_A}$ for Gaussian modulation.

\begin{table}[!htb]
\centering
\caption{Excess Noise in Daylight~\cite{Kish2020}. The values are normalised to shot noise and are expressed in Shot Noise Units (SNU).}
\label{ExcessNoise}
\tabcolsep=0.05cm
\scalebox{0.94}{
\begin{tabular}{!{\vrule width \heavyrulewidth}l!{\vrule width \heavyrulewidth}ll!{\vrule width \heavyrulewidth}}
\bottomrule
$\epsilon$& Source & Value (SNU)                      \\ 
\toprule
\bottomrule
                     & Time-of-arrival fluctuations  & 0.0060 \\
Channel              & Atmospheric relative intensity noise in local oscillator & 0.0100 \\
Excess               & Relative intensity noise in local oscillator & 0.0018 \\
Noise,               & Modulation noise & 0.0005 \\ 
$\epsilon_{\mathrm{ch}}$ & Background noise & 0.0002\\ 
                     & Relative intensity noise in signal & 0.0001 \\ 
\toprule
\bottomrule
Detection            & Electronic noise & 0.0130                       \\
Excess               & Anaogue-to-digital converter noise & 0.0002                       \\
Noise                & Detector overlap & 0.0001                      \\
$\epsilon_{\mathrm{det}}$ & Local oscillator subtraction noise & 0.0001                   \\
                     & Local oscillator to signal leakage & 0.0001                       \\ 
\toprule
\multicolumn{1}{l}{} &  & \multicolumn{1}{l}{} 
\end{tabular}}
\vspace{-20mm}
\end{table}

\subsection{$M$-PSK DM-CVQKD}
\begin{figure}[h]
     \centering
     \begin{subfigure}[b]{0.155\textwidth}
         \centering
         \includegraphics[width=\textwidth]{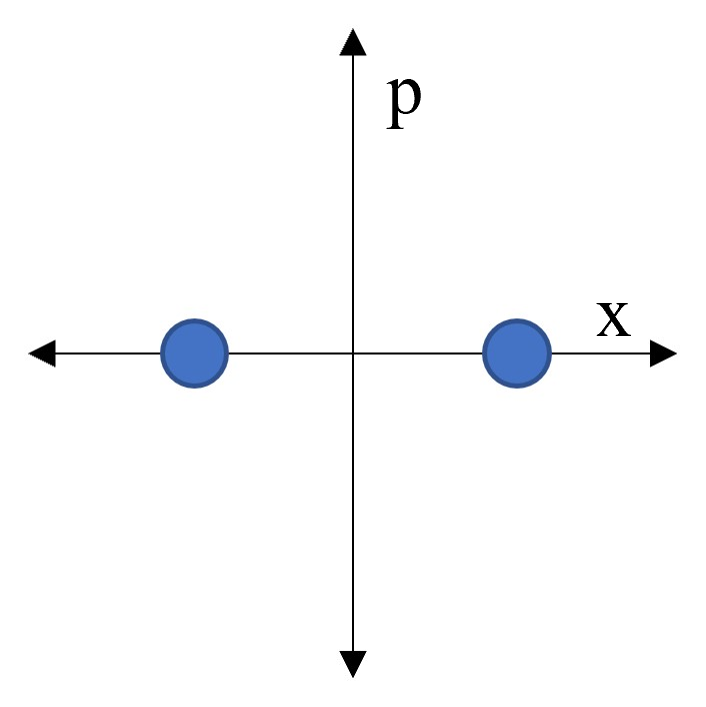}
         \caption{2-PSK}
         \label{fig:2state}
     \end{subfigure}
     \hfill
     \begin{subfigure}[b]{0.155\textwidth}
         \centering
         \includegraphics[width=\textwidth]{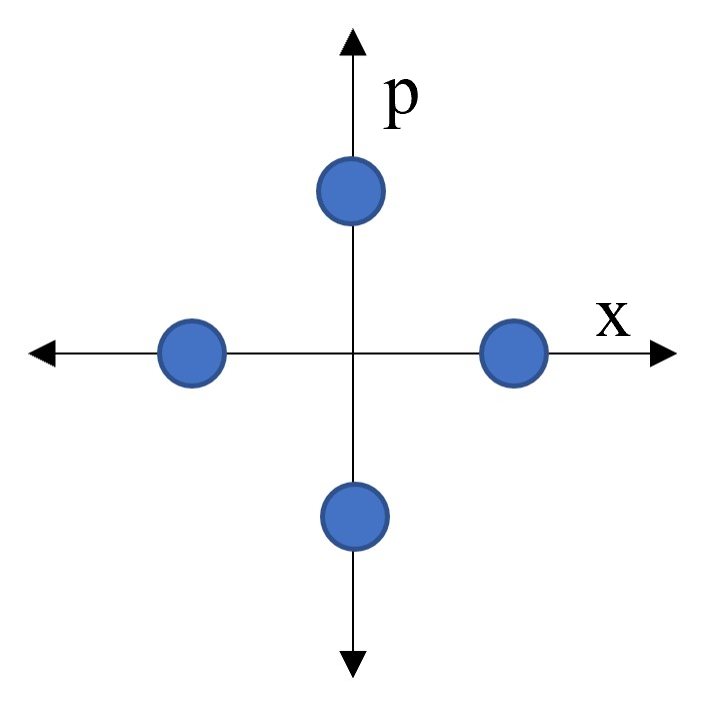}
         \caption{4-PSK}
         \label{fig:4state}
     \end{subfigure}
     \hfill
     \begin{subfigure}[b]{0.155\textwidth}
         \centering
         \includegraphics[width=\textwidth]{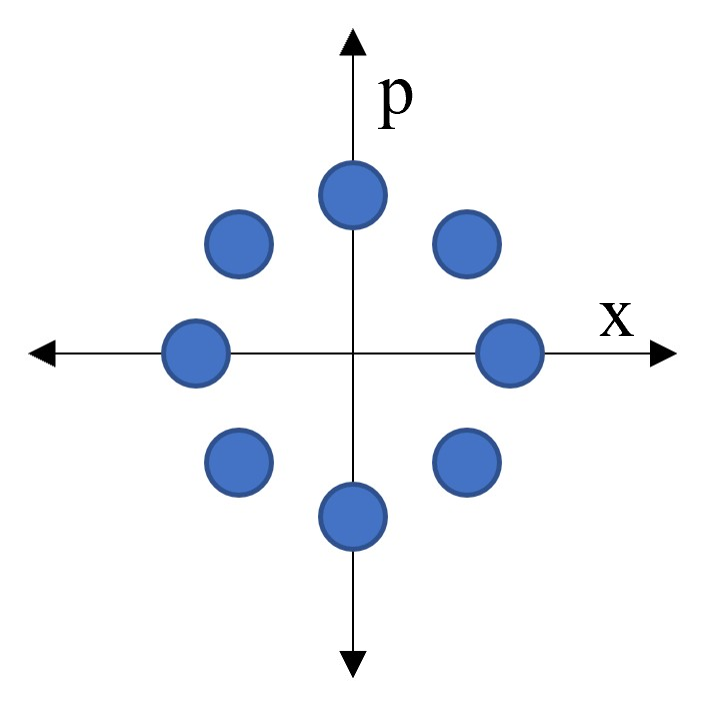}
         \caption{8-PSK}
         \label{fig:8state}
     \end{subfigure}
     \caption{Constellation diagrams of the 2,4,8-PSK protocols on the optical phase space. The coherent states have been modulated with a constant $\alpha$ and have equal probability. x = amplitude quadrature, p = phase quadrature.}
     \label{fig:DMCVQKDConstellationDiagrams}
\end{figure}

In the 2-PSK protocol, Alice sends one of two coherent states with equal probability (0.5). In the 4-PSK protocol, Alice sends one of four coherent states with equal probability (0.25). In the 8-PSK protocol, Alice sends one of eight coherent states with equal probability (0.125). This is summarised below~\cite{Zhao2020}: 
\begin{itemize}
    \item $\mathcal{S}_2 = \{\ket{\alpha}, \ket{\alpha e^{i\pi}}\}$
    \item $\mathcal{S}_4 = \{\ket{\alpha}, \ket{\alpha e^{i\pi/2}},\ket{\alpha e^{i\pi}}, \ket{\alpha e^{i3\pi/2}}\}$
    \item $\mathcal{S}_8 = \{\ket{\alpha}, \ket{\alpha e^{i\pi/4}},\ket{\alpha e^{i\pi/2}}, \ket{\alpha e^{i3\pi/4}}, \ket{e^{i\pi}}, \ket{e^{i5\pi/4}}, \ket{e^{i3\pi/2}}, \ket{e^{i7\pi/8}}\}$
\end{itemize}
The $M$-PSK ($M$ = 2, 4, 8) protocols use coherent states with a magnitude $\alpha$, and are represented in the optical phase space as shown in Figure \ref{fig:DMCVQKDConstellationDiagrams}.

The DM-CVQKD covariance matrix which describes the discretely modulated coherent state sent from Alice to Bob, with security under collective attacks, has the same form as the Gaussian modulation scheme~\cite{Zhang2012},
\begin{equation}
    \label{covariancematrix_DMCVQKD}
    \gamma_{AB} = 
    \begin{bmatrix} (V_A + 1)\boldsymbol{\mathrm{I}} & \sqrt{T}Z_M\boldsymbol{\sigma_z} \\ \sqrt{T}Z_M\boldsymbol{\sigma_z} &  T(V_A + 1 + \chi_{\mathrm{line}})\boldsymbol{\mathrm{I}} \end{bmatrix}.
\end{equation}
In this case, the correlation coefficient, $Z_M$, varies between each $M$-PSK protocol:
\begin{itemize}
    \item $Z_{\mathrm{2}} = \alpha^2(\zeta_0^{3/2}\zeta_1^{-1/2} + \zeta_1^{3/2}\zeta_0^{-1/2})$,
    \item $Z_{\mathrm{4}} = 2\alpha^2\sum_{k=0}^{3}(\zeta_{k-1}^{3/2}\zeta_k^{-1/2})$,
    \item $Z_{\mathrm{8}} = 2\alpha^2\sum_{k=0}^{7}(\zeta_{k-1}^{3/2}\zeta_k^{-1/2})$.
\end{itemize}
The parameter $\zeta_k$ varies for each $M$-PSK protocol:

2-PSK protocol:
\begin{itemize}
  \item $\zeta_0 = e^{-\alpha^2}\mathrm{\cosh{\alpha^2}}$,
  \item $\zeta_1 = e^{-\alpha^2}\mathrm{\sinh{\alpha^2}}$.
\end{itemize}

4-PSK protocol:
\begin{itemize}
  \item $\zeta_{0,2} = \frac{1}{2}e^{-\alpha^2}(\mathrm{\cosh{\alpha^2}} \pm \mathrm{\cos{\alpha^2}})$,
  \item $\zeta_{1,3} = \frac{1}{2}e^{-\alpha^2}(\mathrm{\sinh{\alpha^2}} \pm \mathrm{\sin{\alpha^2}})$.
\end{itemize}

8-PSK protocol:
\begin{itemize}
  \item $\zeta_{0,4} = \frac{1}{4}e^{-\alpha^2}(\mathrm{\cosh{\alpha^2}} + \mathrm{\cos{\alpha^2}} \pm 2\cos{\frac{\alpha^2}{\sqrt{2}}}\cosh{\frac{\alpha^2}{\sqrt{2}}})$,
  \item $\zeta_{1,5} = \frac{1}{4}e^{-\alpha^2}(\mathrm{\sinh{\alpha^2}} + \mathrm{\sin{\alpha^2}} \pm \sqrt{2}\cos{\frac{\alpha^2}{\sqrt{2}}}\sinh{\frac{\alpha^2}{\sqrt{2}}} \pm \sqrt{2}\sin{\frac{\alpha^2}{\sqrt{2}}}\cosh{\frac{\alpha^2}{\sqrt{2}}})$,
  \item $\zeta_{2,6} = \frac{1}{4}e^{-\alpha^2}(\mathrm{\cosh{\alpha^2}} - \mathrm{\cos{\alpha^2}} \pm 2\sin{\frac{\alpha^2}{\sqrt{2}}}\sinh{\frac{\alpha^2}{\sqrt{2}}})$,
  \item $\zeta_{3,7} = \frac{1}{4}e^{-\alpha^2}(\mathrm{\sinh{\alpha^2}} - \mathrm{\sin{\alpha^2}} \mp \sqrt{2}\cos{\frac{\alpha^2}{\sqrt{2}}}\sinh{\frac{\alpha^2}{\sqrt{2}}} \pm \sqrt{2}\sin{\frac{\alpha^2}{\sqrt{2}}}\cosh{\frac{\alpha^2}{\sqrt{2}}})$,
\end{itemize} 
where $\alpha = \sqrt{\frac{V_A}{2}}$.

The asymptotic limit secret key rate is calculated as shown in Equations \ref{SKR}-\ref{Chet_Dhet}. However, $Z_M$ is used for the $Z$ term.

\subsection{$M$-QAM}
In~\cite{Denys2021}, a security proof for the $M$-PSK and $M$-QAM DM-CVQKD protocols without Gaussian optimality under collective attacks from an eavesdropper in the asymptotic limit was developed. Unfortunately, using the security proof with the transmittances and excess noise in a typical satellite-to-ground link (Table \ref{ExcessNoise}), $M$-PSK is not capable of producing a positive SKR. A more attractive protocol is $M$-QAM as it produces positive SKRs at lower transmittances and higher levels of excess noise.
In $M$-QAM, $M$ coherent states are modulated to be distributed equidistantly with each other on the optical phase space. By assigning a non-uniform probability to each coherent state, $M$-QAM can be tailored to a discretised Gaussian distribution to further increase the mutual information between Alice and Bob.

\begin{figure}[!htb]
    \centering
    \begin{subfigure}[b]{0.45\textwidth}
         \centering
         \includegraphics[width=\textwidth]{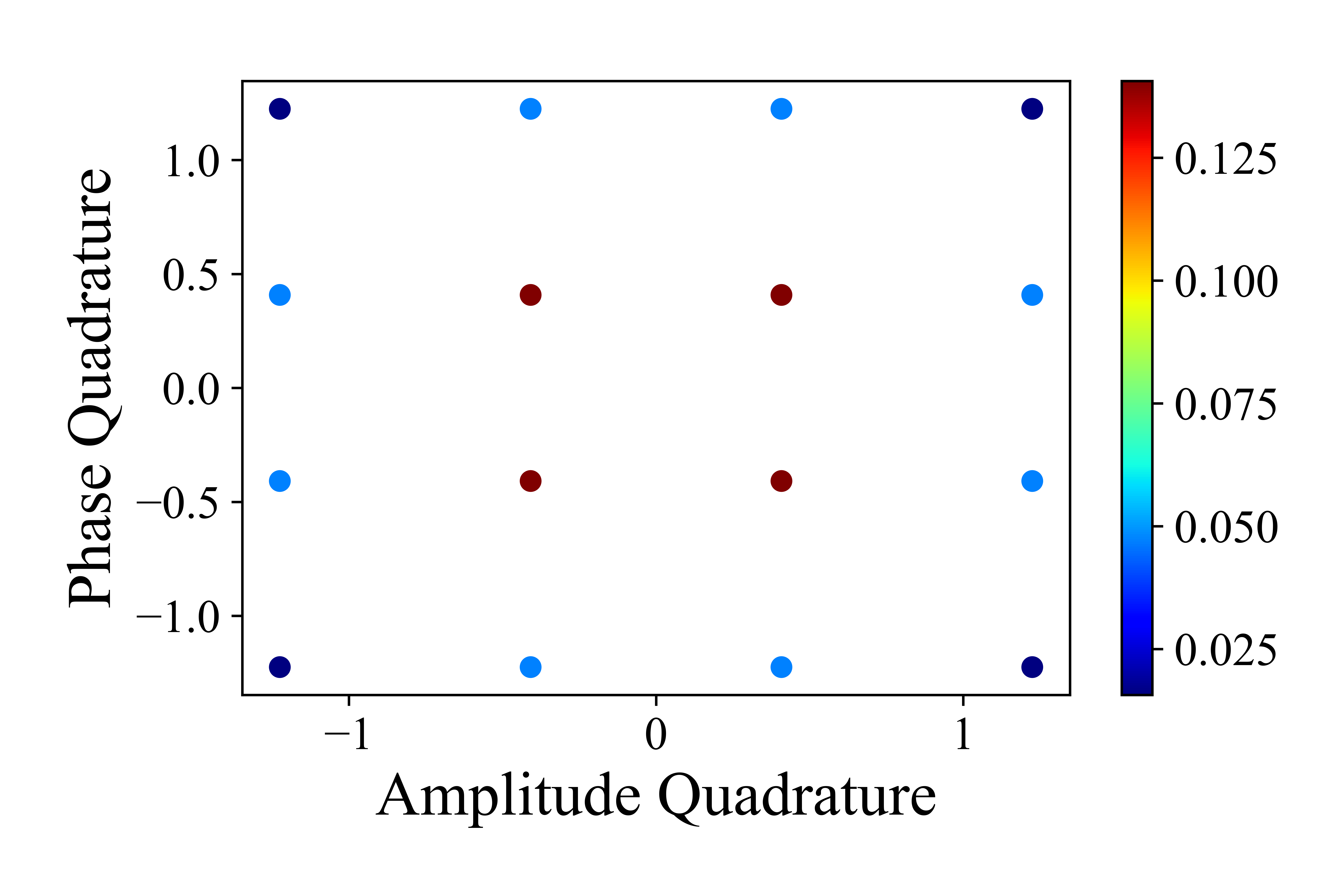}
         \caption{}
         \label{fig:16QAM}
     \end{subfigure}
    \begin{subfigure}[b]{0.45\textwidth}
         \centering
         \includegraphics[width=\textwidth]{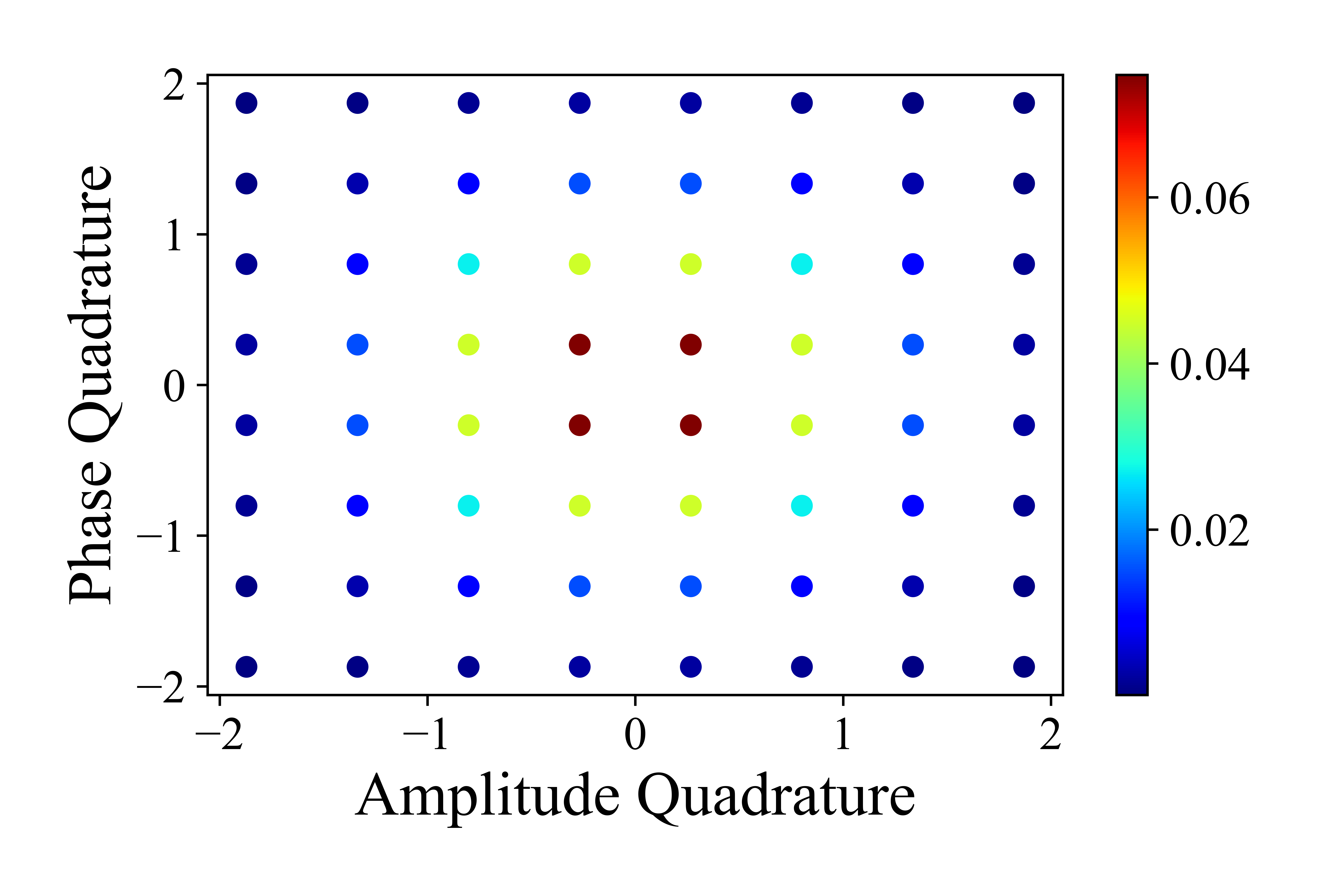}
         \caption{}
         \label{fig:64QAM}
     \end{subfigure}
     \begin{subfigure}[b]{0.45\textwidth}
         \centering
         \includegraphics[width=\textwidth]{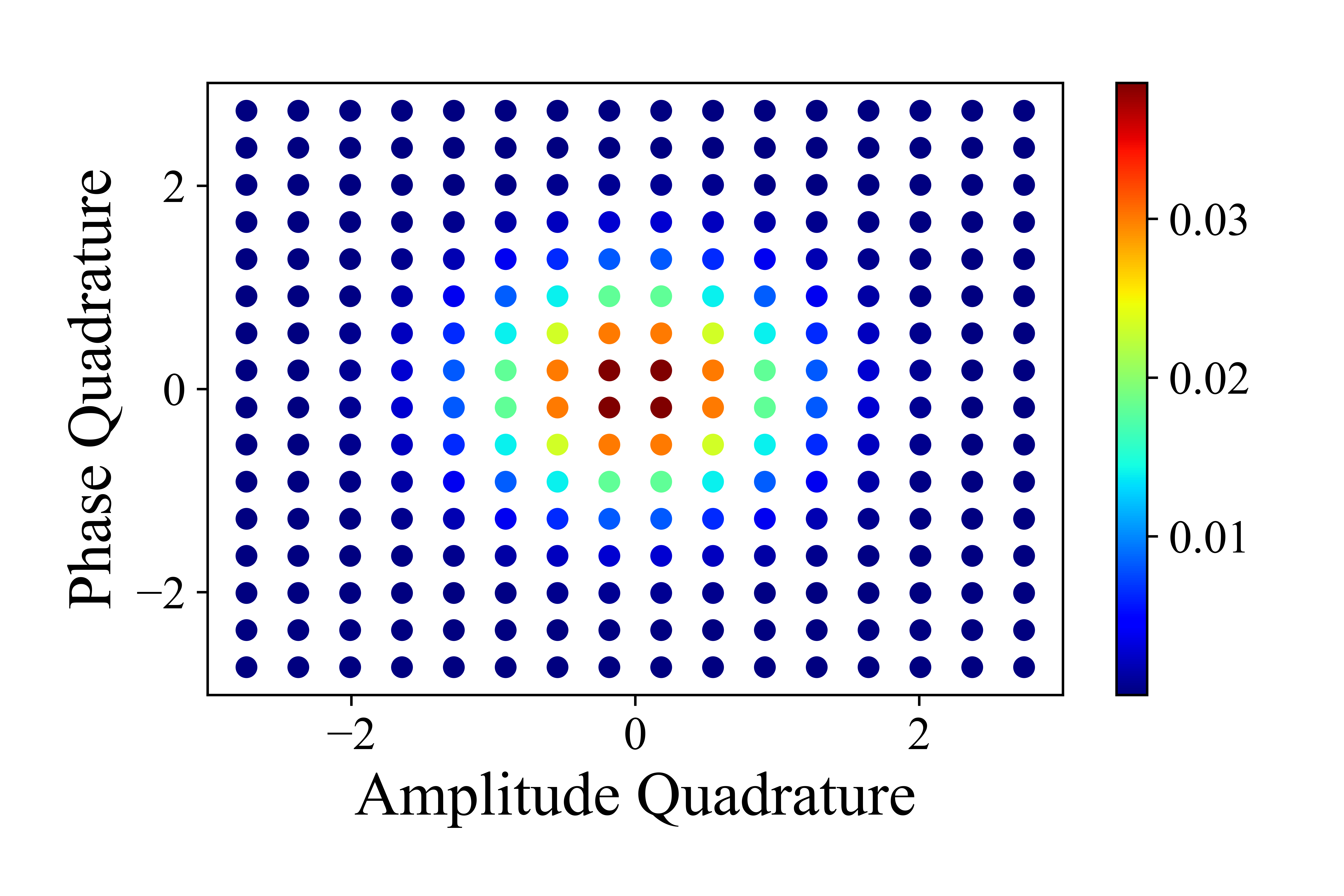}
         \caption{}
         \label{fig:256QAM}
     \end{subfigure}
    \caption{$M$-QAM with probabilities based on the binomial distribution. (a) 16-QAM (b) 64-QAM (c) 256-QAM.}
    \label{fig:QAM}
    \vspace{-5mm}
\end{figure}

The modulated coherent state is described as
\begin{equation}
    \label{alpha_QAM}
    \begin{aligned}
        \alpha_{k, l} = \frac{\alpha \sqrt{2}}{\sqrt{m - 1}}\left(k - \frac{m - 1}{2}\right) + i\frac{\alpha \sqrt{2}}{\sqrt{m - 1}}\left(l - \frac{m - 1}{2}\right),
    \end{aligned}
\end{equation}
where $M = m^2$ and the coherent states are equidistantly spaced between $-\sqrt{m - 1}$ and $\sqrt{m - 1}$ in the phase and amplitude quadratures. Here, $k = l = {0, 1, ..., (m - 1)}$. The probability of each coherent state, $p_{k, l}$, can follow either a binomial distribution,
\begin{equation}
    \label{probability_Binomial}
    \begin{aligned}
        p_{k, l} = \frac{1}{2^{2(m - 1)}}\binom{m - 1}{k}\binom{m - 1}{l},
    \end{aligned}
\end{equation}
or a discrete Gaussian distribution,
\begin{equation}
    \label{probability_discreteGaussian}
    \begin{aligned}
        p_{k, l} \sim \exp(-v\left(x^2 + p^2\right)),
    \end{aligned}
\end{equation}
where $x = \frac{\alpha \sqrt{2}}{\sqrt{m - 1}}\left(k - \frac{m - 1}{2}\right)$ and $p = \frac{\alpha \sqrt{2}}{\sqrt{m - 1}}\left(l - \frac{m - 1}{2}\right)$. Here, $v$ is a free parameter which is optimised to maximise the SKR. For example, Figure \ref{fig:QAM} shows 16-QAM, 64-QAM, and 256-QAM on the optical phase space with the probability of each modulated coherent state based on the binomial distribution. It can be seen that a greater number of coherent states better approximates the Gaussian distribution.

The asymptotic limit SKR is calculated as in Equation \ref{SKR}. However, the mutual information is calculated as
\begin{equation}
    \label{mutual_information_DBL}
    \begin{aligned}
        I_{AB, \mathrm{hom}} &= \frac{1}{2}\log_2{(1 + \frac{TV_A}{2 + T\epsilon})}, \\
        I_{AB, \mathrm{het}} &= \log_2{(1 + \frac{TV_A}{2 + T\epsilon})}.
    \end{aligned}
\end{equation}
The Holevo information is determined from the covariance matrix given by,
\begin{equation}
    \label{covariancematrix_DBL}
    \Gamma_{AB}^* = 
    \begin{bmatrix} \left(V_A + 1\right)\boldsymbol{\mathrm{I}} & Z^*{\sigma_z} \\ Z^*{\sigma_z} &  \left(1 + TV_A + T\epsilon\right)\boldsymbol{\mathrm{I}} \end{bmatrix}.
\end{equation}
The Holevo information is therefore,
\begin{equation}
    S_{BE} = G\left(\frac{\lambda_1 - 1}{2}\right) + G\left(\frac{\lambda_2 - 1}{2}\right) - G\left(\frac{\lambda_3 - 1}{2}\right),
\end{equation}
where $\lambda_1$ and $\lambda_2$ are the symplectic eigenvalues of the covariance matrix, $\Gamma_{AB}^*$. The symplectic eigenvalue, $\lambda_3$, is calculated as
\begin{equation}
    \begin{aligned}
        \lambda_{\mathrm{3, hom}} &= \sqrt{\left(V_A + 1\right)\left(V_A + 1 - \frac{Z^{*2}}{1 + TV_A + T\epsilon}\right)}, \; \mathrm{and} \\
        \lambda_{\mathrm{3, het}} &= V_A + 1 - \frac{Z^{*2}}{2 + TV_A + T\epsilon},
    \end{aligned}
\end{equation}
for homodyne and heterodyne detection, respectively.
For an arbitrary modulation, the lower bound of the correlation coefficient, $Z^*$, can be calculated as 
\begin{equation}
    \begin{aligned}
        Z^*(T, \epsilon) = 2\sqrt{T}\Tr(\tau^{\frac{1}{2}}\opan{a}\tau^{\frac{1}{2}}\opdag{a}) - \sqrt{2T\epsilon w},
    \end{aligned}
\end{equation}
where $\opan{a}$ and $\opdag{a}$ are the annihilation and creation operators, respectively, $\epsilon$ is the total excess noise, and $\tau$ is the density matrix of the modulation:
\begin{equation}
    \begin{aligned}
        \tau = \sum_k p_k \ket{\alpha_k} \bra{\alpha_k} 
    \end{aligned}.
\end{equation}
$w$ is defined as
\begin{equation}
    \begin{aligned}
        w =  \sum_{k} p_k \left( \ket{\alpha_k}\opdag{a}_{\tau}\opan{a}_{\tau}\ket{\alpha_k} - \left|\bra{\alpha_k}\opan{a}_{\tau}\ket{\alpha_k} \right|^2\right) 
    \end{aligned}.
\end{equation}

Although $M$-QAM is a DM-CVQKD protocol, the assignment of a probability following a Gaussian distribution to each modulated coherent state raises the question of whether or not it loses its DM-CVQKD features, especially when $M$ is large. In particular, the distinguishability of the coherent states on the optical phase space. In addition, the calculation of SKRs in the finite size limit for $M$-QAM remains an open question \cite{Denys2021}. As a result, only the SKRs in the asymptotic limit will be studied in a satellite-to-ground context.


\section {Finite size effects}
The SKR calculation (Equation \ref{SKR}-\ref{Chet_Dhet}) is for the idealistic asymptotic limit in which an infinite number of symbols are sent between Alice and Bob, and provides and upper bound to the achievable SKR. The finite size limit SKR, which represents a realistic case in which a finite number of symbols are sent between Alice and Bob~\cite{Mani2021}, can be calculated as 
\begin{equation}
    \label{F_SKR_wrong}
    \begin{aligned} \mathrm{SKR}_{\mathrm{fin}} = f(1 - \mathrm{FER})(1 - v)[\beta I_{AB} - S_{BE} - \delta n_{\mathrm{privacy}}],
    \end{aligned}
\end{equation}
where $f$ is the laser repetition rate, FER is the frame error rate, $v$ is the fraction of symbols excluded for channel parameter estimation, and $\delta n_{\mathrm{privacy}}$ represents the proportion of the key further attributed to information gained by the eavesdropper to reflect the validity of estimated channel parameters in determining the Holevo information. However, to account for high error rate error correcting codes in long-distance CVQKD, the finite size limit SKR must be amended to \cite{Johnson2017, Jeong2022}:
\begin{equation}
    \label{F_SKR}
    \begin{aligned} \mathrm{SKR}_{\mathrm{fin}} = f[(1 - \mathrm{FER})\beta I_{AB} - S_{BE} - \delta n_{\mathrm{privacy}}].
    \end{aligned}
\end{equation}
It should be noted that the term $(1-v)$ has been excluded as shown in composable security proofs where Alice can estimate channel parameters after error correction \cite{Ghorai2019, Hosseinidehaj2021}. Here, $\delta n_{\mathrm{privacy}}$ is calculated as 
\begin{equation}
    \label{privacy}
    \begin{aligned}
         \delta n_{privacy} = \frac{(d + 1)^2}{\sqrt{N}} +\frac{ 4(d+1)\sqrt{\log_2(\frac{2}{\epsilon_s})}}{\sqrt{N}} + \frac{2\log_2(\frac{2}{\epsilon^2 \epsilon_s})}{\sqrt{N}} + \frac{\frac{4\epsilon_s d}{\epsilon \sqrt{N}}}{\sqrt{N}},
    \end{aligned}
\end{equation}
where $d$ is a discretisation parameter, $\epsilon_s$ is a smoothing parameter, $\epsilon$ is a security parameter representing the probability that the key is not secret, and $N$ is the total number of symbols sent between Alice and Bob, the details of which can be found in previous work \cite{Kish2020, Hosseinidehaj2020, Leverrier2015, Lupo2018}.


Reconciliation methods can be split into Multidimensional (MD) reconciliation and Multilevel Coding and Multistage Decoding (MLC-MSD) reconciliation~\cite{Mani2021}. MLC-MSD reconciliation is generally employed for CVQKD at SNRs greater than 0 dB, while MD reconciliation is employed for CVQKD at SNRs less than 0 dB. Low-density parity-check (LDPC) codes can be used with MD and MLC-MSD reconciliation. Multiedge-type LDPC (MET-LDPC) codes are regarded as suitable for both MD and MLC-MSD reconciliation. Recent practical MD reconciliation efficiencies and FER asymptotically approach 100\% as the SNR decreases~\cite{Mani2021}. 

The dependency of $\beta$ and FER on the SNR in (Equation \ref{Beta_FER_SNR}) have been empirically determined from \cite{Mani2021}, and are used to calculate and analyse the satellite-to-ground SKRs in the finite size limit. The FER equation is for $N = 10^6$. However, the general trend shows that the FER decreases as $N$ increases \cite{Mani2021}. As a result, a code block length of $N = 10^{11}$ has been used for finite size limit SKR calculations. The coefficients, $c_i$, for $\beta$ are displayed in Table \ref{beta}. The calculation of the FER uses the coefficients $m_1 = 0.8218$, $m_2 = -19.46$, and $m_3 = -298.1$ as shown in Equation \ref{Beta_FER_SNR}.

\begin{equation}
    \label{Beta_FER_SNR}
        \begin{aligned}
            \beta_{\mathrm{MLC-MSD/MD}} &= c_1^{c_2\mathrm{SNR}} - c_3^{c_4\mathrm{SNR}},\\
            \mathrm{FER} &= \frac{1}{2}(1 + m_1\arctan(m_2\mathrm{SNR} + m_3)).
        \end{aligned}
\end{equation}
The SNR is calculated as 
\begin{equation}
    \label{SNR}
        \begin{aligned}
            \mathrm{SNR} = 10\log_{10}{\left(\frac{T|\alpha|^2}{|\alpha|^2 + (1-T)\chi_{\mathrm{tot}}}\right)} ,
        \end{aligned}
\end{equation}
and depends on $\alpha$, transmittance ($T$) and excess noise ($\chi_{\mathrm{tot}}$) between Alice and Bob. Note that $\beta_{\mathrm{MLC-MSD}}$, $\beta_{\mathrm{MD}}$, and FER in Equation \ref{Beta_FER_SNR} are only valid when they have a value between 0 and 1.
\begin{table}[!htb]
    \caption{Coefficients of $\beta$.}
    \centering
    \label{beta}
    \tabcolsep=0.25cm
    \scalebox{1}{
    \begin{tabular}{!{\vrule width \heavyrulewidth}l!{\vrule width \heavyrulewidth}ll!{\vrule width \heavyrulewidth}} 
    \bottomrule
    Coefficient & MLC-MSD & MD                     \\ 
    \toprule
    \bottomrule
    $c_1$                     & 0.9655  & -0.0825 \\
    $c_2$                     & 0.0001507 & 0.1834 \\
    $c_3$                     & -0.04696 & 0.9821 \\
    $c_4$                     & -0.2238 & -0.00002815 \\ 
    \toprule
    \multicolumn{1}{l}{} &  & \multicolumn{1}{l}{} 
    \end{tabular}}
    \vspace{-15mm}
\end{table}

\section{Satellite-to-Ground Channel Model}
Clouds, atmospheric turbulence, and atmospheric aerosols are the three main sources of signal degradation in the atmosphere that cause the transmittance of the signal to decrease through attenuation. The model developed considers the following: geometric losses due to the distance between Alice (satellite) and Bob (OGS), the hardware used, scintillation losses due to the atmospheric turbulence, scattering losses due to atmospheric aerosols. Clouds are not included in the model as they effectively destroy the signal and act as a blockage, completely attenuating the signal. The solution to this is accurate cloud coverage analyses and OGS network site diversity~\cite{Bennet2020} to spatiotemporally maximise channel links between Alice and Bob regardless of environmental conditions. In addition, weather conditions such as rain, snow, and hail effectively destroy the signal and can cause damage to an OGS telescope~\cite{Muhammad2005}. Therefore, in the presence of clouds, rain, snow, and hail, satellite-to-ground CVQKD is not ideal.

Atmospheric losses depend on the thickness of the atmosphere or the atmospheric mass the signal propagates through. On average, 95\% of the total atmosphere mass is within the first 20~km from ground to zenith~\cite{Liorni2019, Zuo2020} and so atmospheric losses can be modelled as approximately confined to this range. 

\begin{figure}[!htb]
    \centering
    \includegraphics[width=0.60\textwidth]{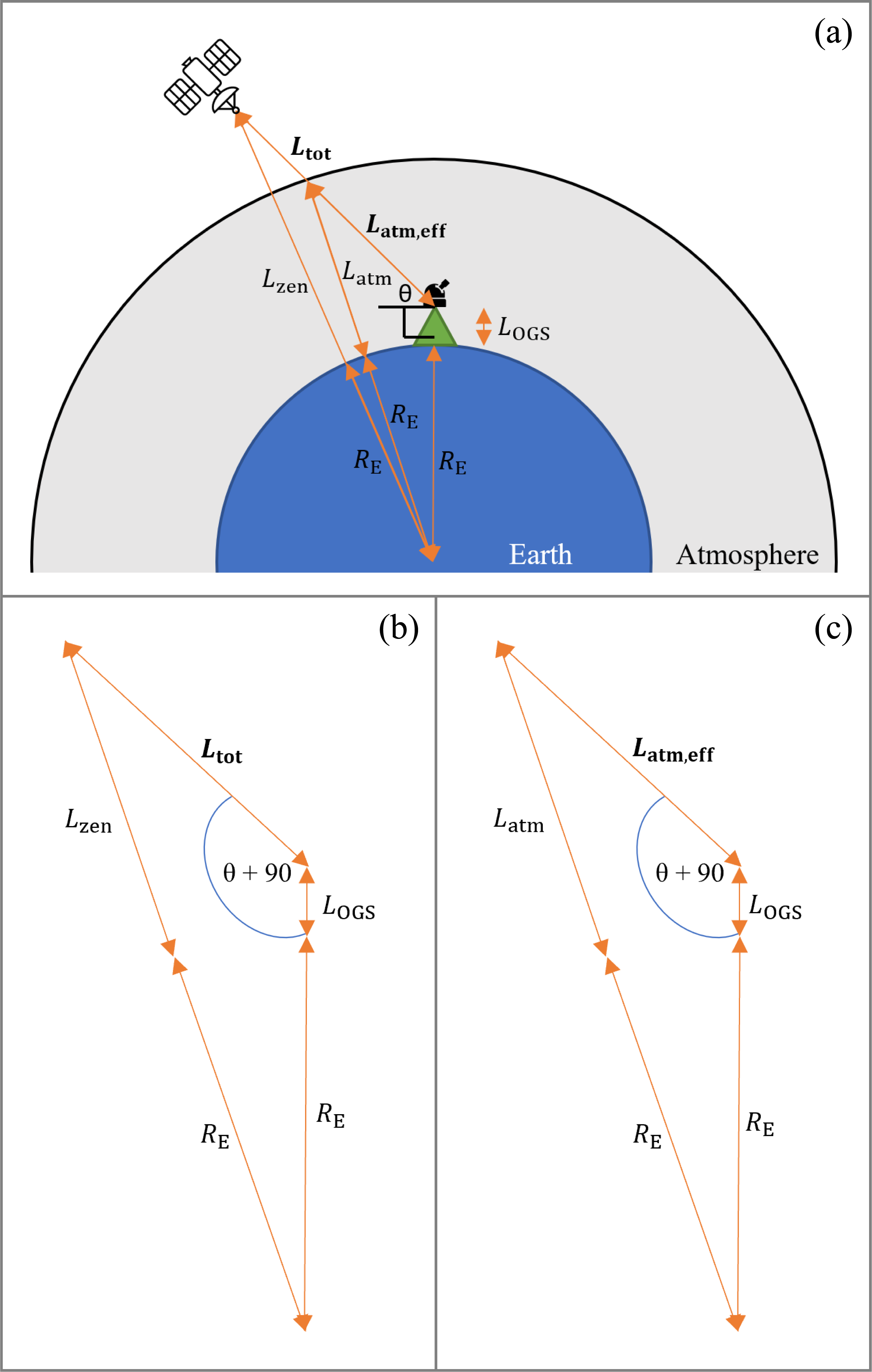}
    \caption{The satellite-to-ground channel model is shown in (a). (b) and (c) shows the trigonometric determination of the total link distance and effective atmosphere thickness, respectively.}
    \label{fig:channel_model}
\end{figure}

Figure \ref{fig:channel_model} displays the channel link between the satellite and OGS where the total link distance and effective atmosphere thickness is a function of the elevation angle. The model assumes a uniform atmosphere thickness of 20~km. The total link distance, $L_{\mathrm{tot}}$, and effective atmosphere thickness, $L_{\mathrm{atm,eff}}$, can be calculated as
\begin{equation}
\label{LinkDistances}
    \begin{aligned}
        L_{\mathrm{tot}} = & \left(R_{\mathrm{E}} + L_{\mathrm{zen}}\right)^2 + \left(R_{\mathrm{E}} + L_{\mathrm{OGS}}\right)^2 -2(R_{\mathrm{E}} + L_{\mathrm{zen}})\left(R_{\mathrm{E}} + L_{\mathrm{OGS}})\cos{(\alpha_1)}\right)^\frac{1}{2} \\
        \alpha_1 = & \arcsin{\left[\cos{(\theta)}\frac{(R_{\mathrm{E}} + L_{\mathrm{OGS}})}{R_{\mathrm{E}} + L_{\mathrm{zen}}}\right]} + (90 - \theta), \; \mathrm{and}
        \\
        L_{\mathrm{atm,eff}} = & (R_{\mathrm{E}} + L_{\mathrm{atm}})^2 + (R_{\mathrm{E}} + L_{\mathrm{OGS}})^2 -2(R_{\mathrm{E}} + L_{\mathrm{atm}})(R_{\mathrm{E}} + L_{\mathrm{OGS}})\cos{(\alpha_2)})^\frac{1}{2} \\
        \alpha_2 = & \arcsin{\left[\cos{(\theta)}\frac{(R_{\mathrm{E}} + L_{\mathrm{OGS}})}{R_{\mathrm{E}} + L_{\mathrm{atm}}}\right]} + (90 - \theta),
    \end{aligned}
\end{equation}
where $R_{\mathrm{E}}$ is the radius of Earth, $L_{\mathrm{OGS}}$ is the altitude of the OGS, $L_{\mathrm{zen}}$ is the altitude of the satellite at zenith (90\degree$\,$ elevation angle), $L_{\mathrm{atm}}$ is the atmospheric thickness containing 95\% of atmospheric mass, and $\theta > 0$ is the elevation angle.

\subsection{Geometric Losses}
Geometric losses, $A_{geo}$ (in units of decibels), arise from the link distance and optical hardware used in the transmission and reception of the signal, and can be estimated from~\cite{Aspelmeyer2003},
\begin{equation}
    \label{losses_geo}
    \begin{aligned}
        A_{\mathrm{geo}} = 10\log_{10}\left(\frac{L_{\mathrm{tot}}^2\lambda^2}{D_{\mathrm{t}}^2 D_{\mathrm{r}}^2} \frac{1}{T_{\mathrm{t}}(1 - L_{\mathrm{p}})T_{\mathrm{r}}}\right) [\mathrm{dB}],
    \end{aligned}
\end{equation}
which depends on the total link distance ($L_{\mathrm{tot}}$), wavelength ($\lambda$), transmitter and receiver aperture diameter ($D_{\mathrm{t}}, D_{\mathrm{r}}$), transmitter and receiver efficiencies ($T_{\mathrm{t}}, T_{\mathrm{r}}$), and pointing loss ($L_{\mathrm{p}}$) which is attributed to the inefficiency of the acquisition, pointing, tracking (APT) system of the OGS as well as beam wandering. Equation \ref{losses_geo} assumes that the receiver is in the far field of the transmitter ($L_{\mathrm{tot}} \geq \frac{D_rD_t}{\lambda}$), the transmitter is diffraction limited, and there is no attenuation from the atmosphere which enables the later addition of other atmospheric losses (scattering and scintillation losses) in the overall model.
\newpage
\subsection{Scattering Losses}
Atmospheric aerosols attenuate the signal through three regimes of scattering~\cite{Kim2001}: Rayleigh scattering caused by air molecules, Mie scattering caused by haze and fog, and geometrical scattering caused by rain, snow etc. For the wavelength used in DM-CVQKD, 1550~nm, the effects of Rayleigh scattering on transmittance is negligible~\cite{Kim2001} and geometrical scattering effects are neglected as satellite-to-ground DM-CVQKD is not ideal during weather conditions such as rain, snow, and hail. 

The Kruse and Kim model~\cite{Grabner2010} is used to model losses related to Mie scattering. The model depends on the wavelength of the signal and the atmospheric visibility, $V$ (Equation \ref{scattering}). The resulting loss per kilometre is then multiplied by the effective atmosphere thickness (Equation \ref{LinkDistances}):
\begin{equation}
    \label{scattering}
    \begin{aligned}
        A_{\mathrm{scat}} = 10\log_{\mathrm{10}}(e)(\frac{3.912}{V})(\frac{\lambda}{550})^{-p} \:\mathrm{[dB/km]}, \\
        p =  
        \begin{cases} 
            1.6 & V \geq 50 \: \mathrm{km} \\
            1.3 & 6 \: \mathrm{km} \leq V < 50 \: \mathrm{km} \\
            0.16V + 0.34 & 1 \:\mathrm{km} \leq V < 6 \:\mathrm{km} \\
            V - 0.5 & 0.5 \:\mathrm{km} \leq V < 1 \:\mathrm{km} \\
            0 & V < 0.5 \:\mathrm{km}.
       \end{cases}
    \end{aligned}
\end{equation}

\subsection{Scintillation Losses}
Atmospheric turbulence causes refractive index variations in the atmosphere and distorts the optical wavefront of the laser, leading to intensity fluctuations and losses in the received signal observed as speckles on an imaging detector~\cite{Bennet2020}. This is a phenomenon known as scintillation, and it can be quantified by the scintillation index, $\sigma_{\mathrm{I}}^2$. As the receiver involves a telescope, it is assumed to have an aperture diameter larger than the irradiance correlation width (lateral intensity coherence length) of the speckles. Therefore, aperture averaging is utilised, decreasing the adverse effects of scintillation~\cite{Giggenbach2008}.

The losses related to scintillation with aperture averaging are described in Equation \ref{scintillation} which depends on the scintillation index, $\sigma_{\mathrm{I}}^2$, and the probability that the received power is below the minimum required power to register a signal, $p_{\mathrm{thr}}$. $p_{\mathrm{thr}}$ is equivalent to the fraction of link outage time. Here, it is assumed that the received signal is a spherical wave. The resulting loss as a function of the scintillation index is given by \cite{Giggenbach2008},
\begin{equation}
    \label{scintillation}
        \begin{aligned}
            A_{sci} = & 4.343\Big(\mathrm{erf}^{-1} \left(2p_{\mathrm{thr}} - 1\right) \left[2\ln(\sigma_I^{2} + 1)\right]^\frac{1}{2} \\
            & -\frac{1}{2}\ln(\sigma_I^{2} + 1)\Big) [\mathrm{dB}].
        \end{aligned}
\end{equation}
The scintillation index can be calculated using Equation \ref{scintillation_index} which depends on the Rytov variance ($\sigma_{\mathrm{R}}^2$), the effective atmosphere thickness ($L_{\mathrm{atm,eff}}$), and the refractive index structure parameter ($C_n^2$). Here. $k = \frac{2\pi}{\lambda}$ is the wave number, $d = D_{\mathrm{r}}(\frac{\pi}{2\lambda L_{\mathrm{atm,eff}}})^{\frac{1}{2}}$, $D_{\mathrm{r}}$ is the aperture diameter of the receiver, and $\lambda$ is the wavelength.
\begin{equation}
    \label{scintillation_index}
        \begin{aligned}
            \sigma_{\mathrm{I}}^{2}(D_{\mathrm{r}})= \mathrm{exp}\bigg\{ \frac{0.20\sigma_R^{2}}{[1 + 0.18d^{2} + 0.20(\sigma_R^{2})^{\frac{6}{5}}]^{\frac{7}{6}}} \\
            + \frac{0.21\sigma_R^{2} [1 + 0.24(\sigma_R^{2})^{\frac{6}{5}}]^{-\frac{5}{6}}]}{1 + 0.90d^{2} + 0.21d^{2}(\sigma_R^{2})^{\frac{6}{5}}} \bigg\} - 1, \\
            \sigma_R^{2} = 2.25k^{\frac{7}{6}} \int_{0}^{L_{\mathrm{atm,eff}}} C_n^{2}(z)(L_{\mathrm{atm,eff}}-z)^{\frac{5}{6}}  \, dz. 
        \end{aligned}
\end{equation}

The refractive index structure parameter, $C_n^2 \; [m^{-\frac{2}{3}}]$, defines the intensity of turbulence in the atmosphere. Although this can be calculated more accurately through models such as the Hufnagel-Valley models~\cite{Andrews2009, Villasenor2020}, $C_n^2$ is set to be a constant parameter, where $C_n^2 = 10^{-16}$ corresponds to low turbulence levels, $C_n^2 = 10^{-14}$ medium turbulence levels, and $C_n^2 = 10^{-13}$ high turbulence levels \cite{Muhammad2005}.

\section{Satellite-to-Ground GM-CVQKD and DM-CVQKD SKR Analysis}
In this section, the SKRs of satellite-to-ground GM-CVQKD and DM-CVQKD are estimated with typical daylight excess noise values (Table \ref{ExcessNoise}), and the losses applied to the total link distance and effective atmosphere thickness. The total attenuation based on losses is defined as 
\begin{equation}
    \label{TotalLoss}
        \begin{aligned}
            A_{\mathrm{tot}} = A_{\mathrm{geo}}(L_{\mathrm{tot}}) + A_{\mathrm{scat}}L_{\mathrm{atm,eff}} + A_{\mathrm{sci}}(L_{\mathrm{atm,eff}}) \; [\mathrm{dB}]
        \end{aligned}
\end{equation}
and the transmittance is calculated as the non-logarithmic inverse of $A_{\mathrm{tot}}$. Of the three sources of attenuation, geometric loss dominates due to the relatively large distance between the satellite and OGS. In calculating geometric losses, the assumption that the receiver is in the far field of the transmitter is governed by $L \geq \frac{D_r D_t}{\lambda}$. In situations where this inequality does not hold, the SKR calculation has been omitted. However, the general trend should still apply.

Several parameters were varied for the analysis of the GM-CVQKD and DM-CVQKD protocols under different satellite orbits (Table \ref{Parameters}). The two values for the visibility and the refractive index structure parameter correspond to good atmospheric conditions where the visibility is high and there is low atmospheric turbulence ($V = 200~\mathrm{km}$, $C_n^2 = 10^{-16}$), and bad atmospheric conditions where visibility is low and there is high atmospheric turbulence ($V = 20~\mathrm{km}$, $C_n^2 = 10^{-13}$). Two values of the receiver aperture diameter were also used to observe the changes in achievable SKR. The modulation variance for Gaussian and discrete modulation were chosen to be close to optimal ($V_A = 5 \;\mathrm{SNU}$ for Gaussian modulation, $V_A = 0.5 \;\mathrm{SNU}$ for $M$-PSK, $V_A = 2 \;\mathrm{SNU}$ for $M$-QAM). The SKRs for GM-CVQKD and DM-CVQKD were calculated as a function of the satellite altitude at zenith at different elevation angles (30\degree, 60\degree, 90\degree) in reference to the satellite altitude at zenith. 

\begin{table} [!htb]
\centering
\caption{CVQKD Link Parameters.}
\label{Parameters}
\scalebox{0.85}{
\begin{tabular}{@{}|ll|@{}}
\bottomrule
Parameter &  Value (Unit) \\ \toprule \bottomrule
Detection  & Homodyne ($M$-PSK, GM-CVQKD) \\ 
           & Heterodyne ($M$-QAM) \\
Modulation Variance ($V_A$) &  0.5 SNU ($M$-PSK) \\ 
                            &  2 SNU ($M$-QAM)\\
                            &  5 SNU (Gaussian)\\
Laser repetition rate ($f$) & 50 MHz \\
Discretisation parameter ($d$) & 5 \\
Smoothing parameter ($\epsilon_s$) & $2 \times 10^{-10}$ \\
Security parameter ($\epsilon$) & $1 \times 10^{-9}$ \\
Wavelength ($\lambda$) &  1550 nm\\
Transmitter Aperture Diameter ($D_{\mathrm{t}}$) & 0.3~m\\
Receiver Aperture Diameter ($D_{\mathrm{r}}$) & 1, 2~m  \\
Transmitter, Receiver Optics Efficiency ($T_{\mathrm{t}}, T_{\mathrm{r}}$)  & 0.9, 0.9 \\
Pointing Loss Efficiency / APT Efficiency ($L_{\mathrm{p}}$) & 0.1 \\
OGS Elevation ($L_{\mathrm{OGS}}$) & 0~km  \\
                                & 1.029~km (Mt. John) \\
Atmosphere Thickness (95\% mass) ($L_{\mathrm{atm}}$)  &  20~km \\
Visibility ($V$) & 20~km (Bad) \\
                 & 200~km (Good) \\ 
Refractive Index Structure Parameter ($C_{n}^2$)  & $10^{-13} \:\mathrm{m}^{-\frac{2}{3}}$ ($\mathrm{Bad}$)  \\
                                                  & $10^{-16} \:\mathrm{m}^{-\frac{2}{3}}$ ($\mathrm{Good}$)\\
Probability Threshold ($p_{thr}$) & $10^{-6}$ \\
Code block length  & $10^{11}$ \\ 
/ Total number of symbols sent ($N$) & \\\toprule
\end{tabular}}
\vspace{-3mm}
\end{table}

\subsection{Asymptotic Limit SKRs of GM-CVQKD and DM-CVQKD Protocols}
This section calculates the asymptotic limit SKRs for satellite-to-ground GM-CVQKD and DM-CVQKD under collective attacks from Eve. A reconciliation efficiency of 90\% has been used for asymptotic limit SKR calculations.

\begin{figure}[!htb]
    \centering
    \begin{subfigure}[b]{0.49\textwidth}
         \centering
         \includegraphics[width=\textwidth]{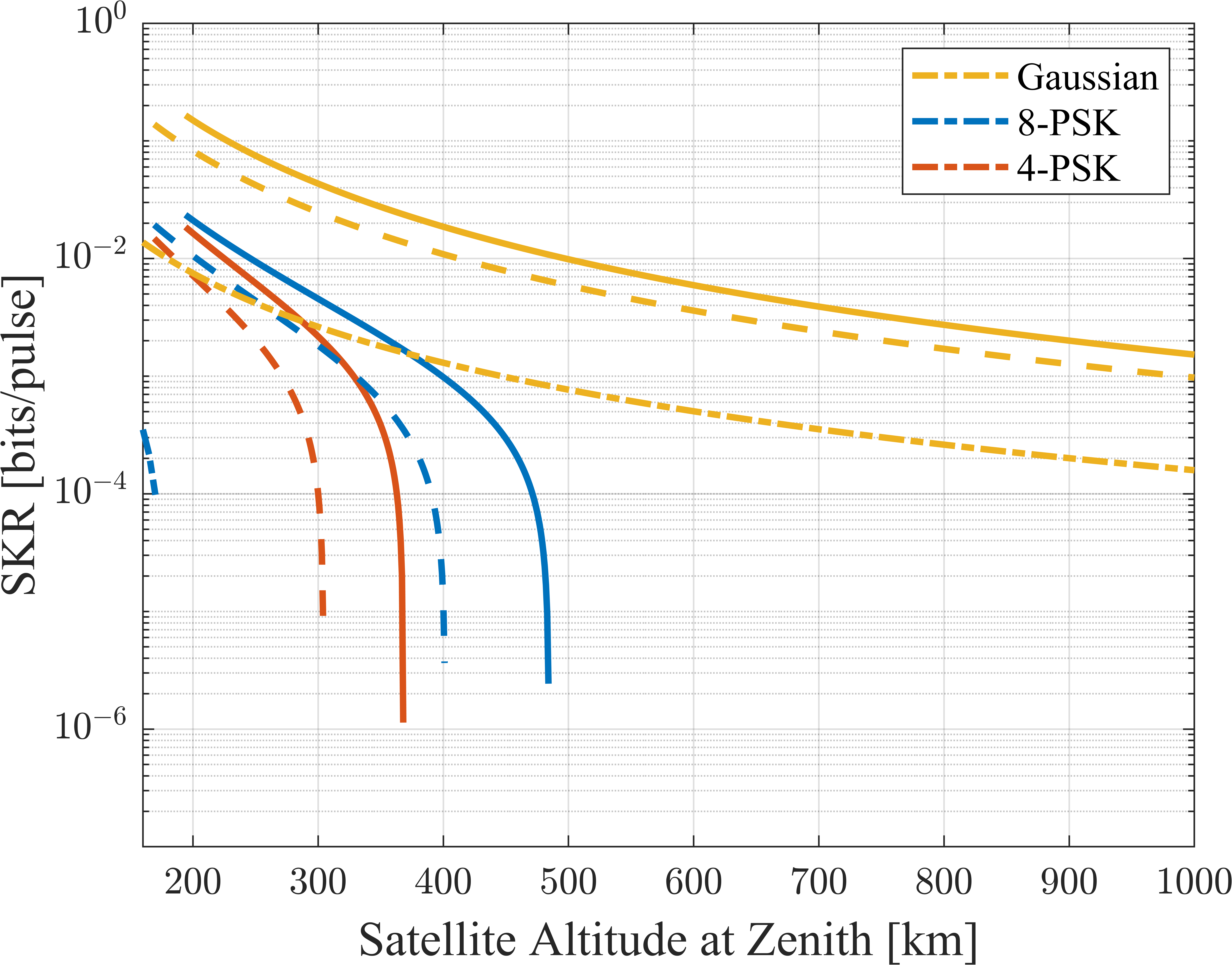}
         \caption{}
         \label{fig:A_good_PSK}
     \end{subfigure}
     \hfill
    \begin{subfigure}[b]{0.48\textwidth}
         \centering
         \includegraphics[width=\textwidth]{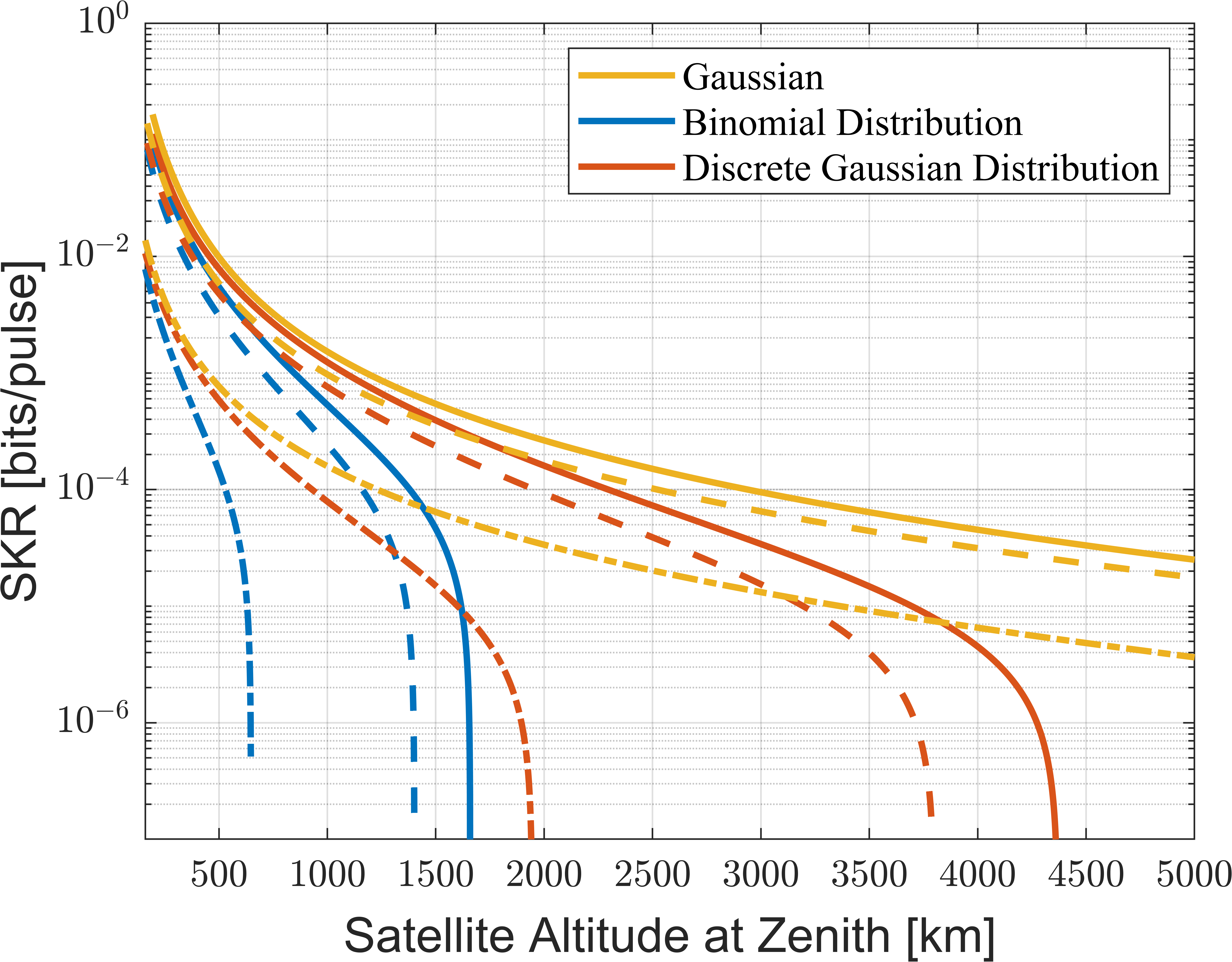}
         \caption{}
         \label{fig:A_good_64QAM}
     \end{subfigure}
     \hfill
     \begin{subfigure}[b]{0.48\textwidth}
         \centering
         \includegraphics[width=\textwidth]{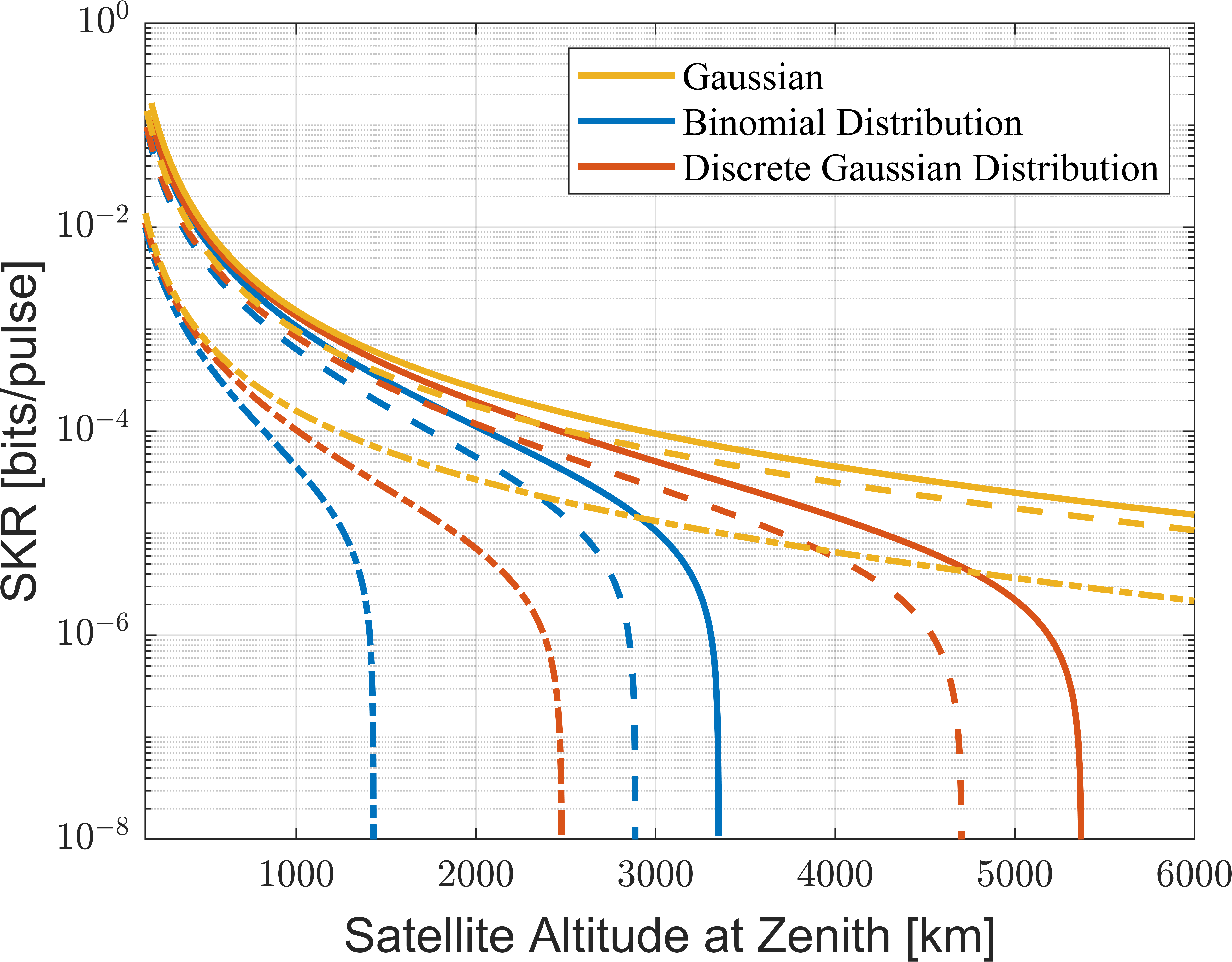}
         \caption{}
         \label{fig:A_good_256QAM}
     \end{subfigure}
     \hfill
    \vspace{-2mm} 
    \caption{Asymptotic limit SKRs as a function of satellite altitude for (a) $M$-PSK, (b) 64-QAM, and (c) 256-QAM DM-CVQKD in relation to GM-CVQKD (yellow) in good atmospheric conditions. The solid lines indicate $\theta =$ 90\degree, dashed lines indicate $\theta =$ 60\degree, dash-dotted lines indicate $\theta =$ 30\degree. $D_r =$ 1~m, $\beta$ = 90\%.}
    \label{fig:A_SKR_Good}
    \vspace{-8mm}
\end{figure}

The results in Figure \ref{fig:A_SKR_Good} show that for a receiver aperture diameter of 1 m and good atmospheric conditions, GM-CVQKD outperforms all DM-CVQKD protocols. The 2-PSK protocol does not produce any positive SKRs and is therefore unsuitable for satellite-to-ground downlinks. As expected, smaller elevation angles produce smaller SKRs as the signal travels through a larger link distance and effective atmosphere thickness, suffering greater attenuation. The $M$-QAM protocol outperforms the $M$-PSK protocol, where $M$-PSK can only achieve positive SKRs at smaller LEO altitudes. The $M$-QAM protocol can produce positive SKRs at small MEOs with the 256-QAM protocol capable of producing positive SKRs at larger link distances.

\begin{figure}[!htb]
    \centering
    \begin{subfigure}[b]{0.49\textwidth}
         \centering
         \includegraphics[width=\textwidth]{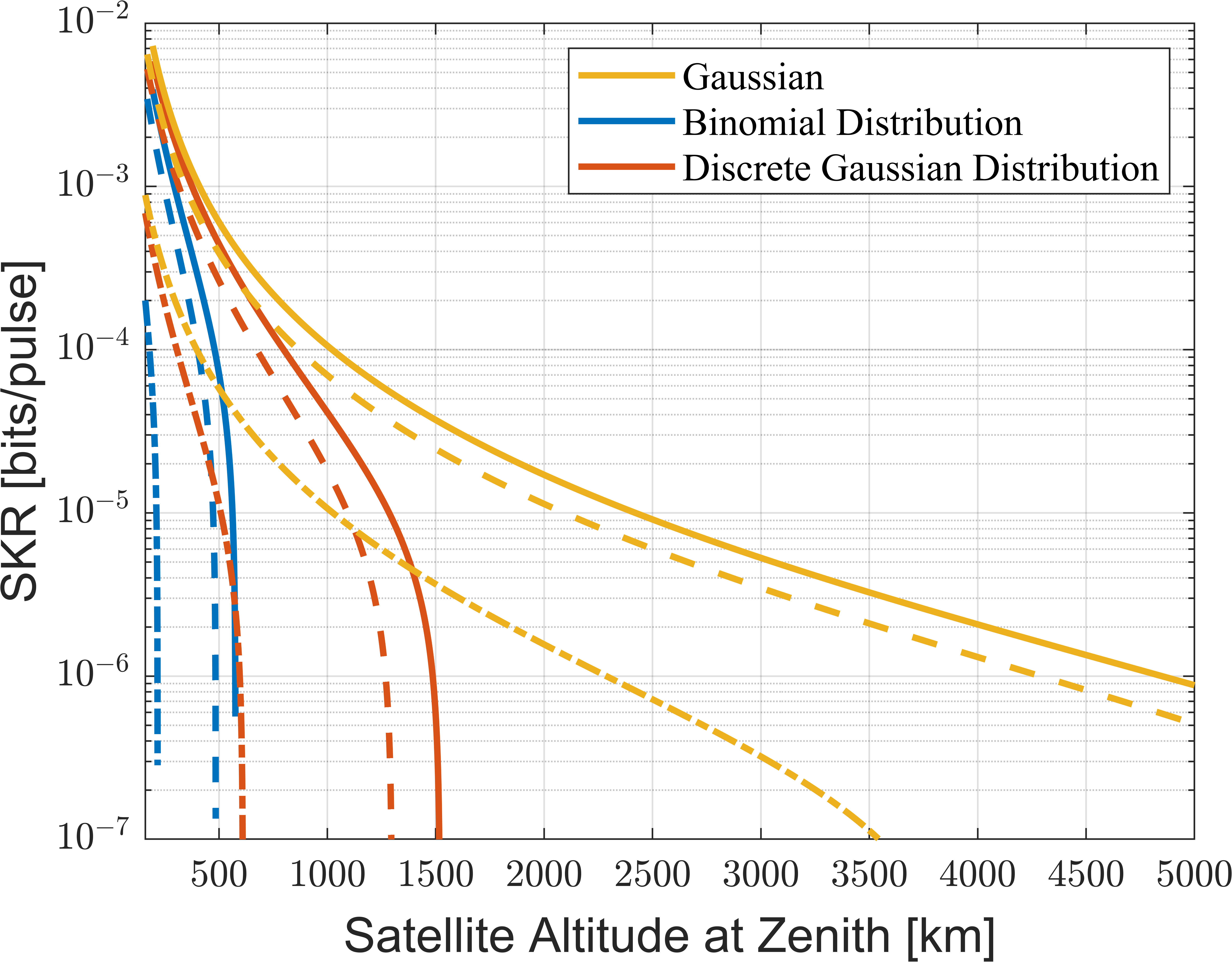}
         \caption{}
         \label{fig:A_bad_64QAM}
     \end{subfigure}
     \hfill
    \begin{subfigure}[b]{0.48\textwidth}
         \centering
         \includegraphics[width=\textwidth]{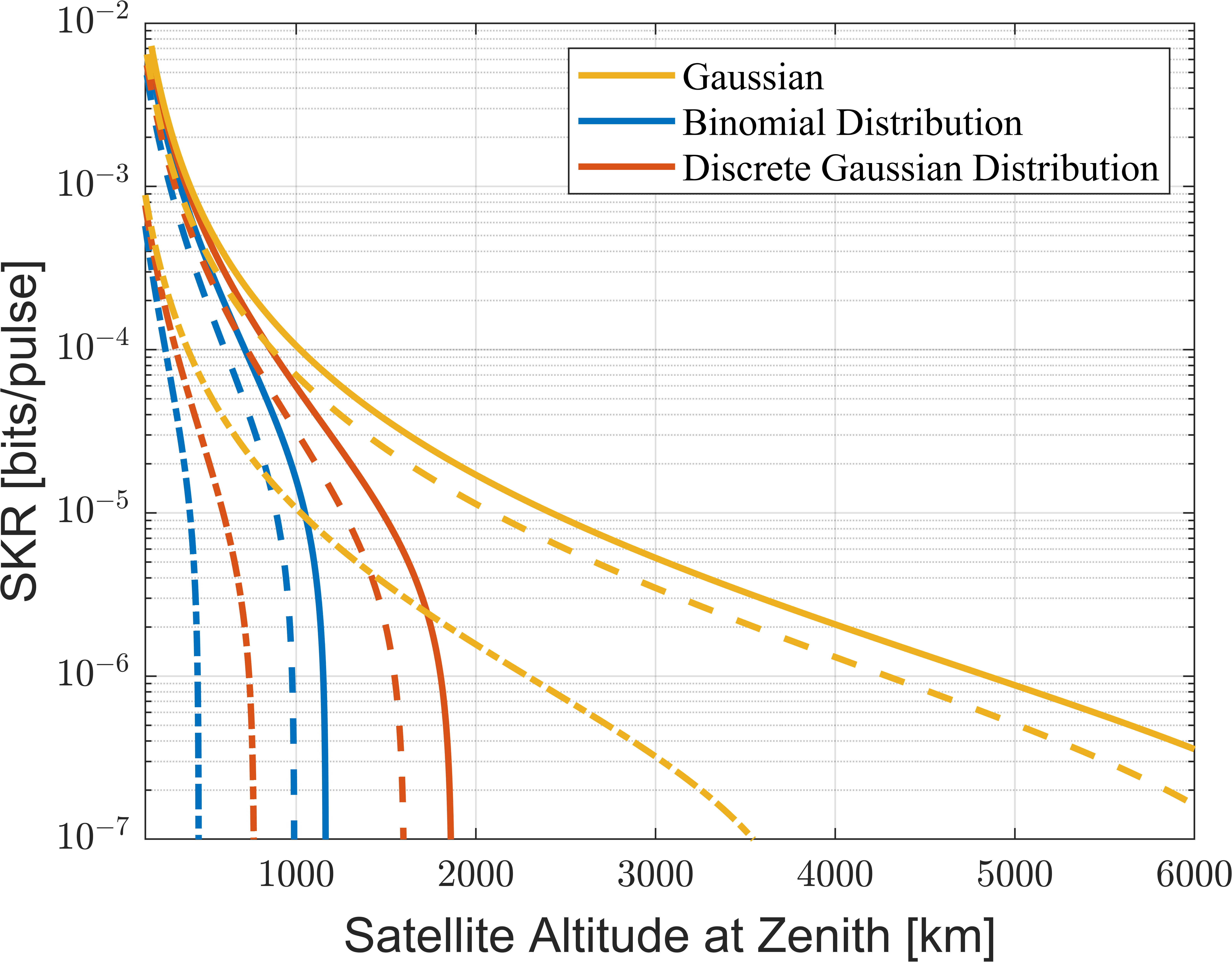}
         \caption{}
         \label{fig:A_bad_256QAM}
     \end{subfigure}
     \hfill
    \vspace{-2mm} 
    \caption{Asymptotic limit SKRs as a function of satellite altitude for (a) 64-QAM and (b) 256-QAM DM-CVQKD in relation to GM-CVQKD (yellow) in bad atmospheric conditions. The solid lines indicate $\theta =$ 90\degree, dashed lines indicate $\theta =$ 60\degree, dash-dotted lines indicate $\theta =$ 30\degree. $D_r =$ 1~m, $\beta$ = 90\%.}
    \label{fig:A_SKR_Bad}
    \vspace{-8mm}
\end{figure}

In bad atmospheric conditions, the $M$-PSK protocol cannot produce positive SKRs in LEO. In addition, the SKRs of GM-CVQKD and $M$-QAM have been significantly decreased (Figure \ref{fig:A_SKR_Bad}).

\subsection{Finite Size Limit SKRs of GM-CVQKD and DM-CVQKD}
As security proofs for DM-CVQKD in the finite size limit are an active field of research \cite{Matsuura2021, Yamano2022, Kanitschar2023}, SKR analysis in the finite size limit was restricted to GM-CVQKD. This analysis uses Equations \ref{F_SKR}-\ref{SNR}. However, it is suspected that the relationship from asymptotic limit to finite size limit DM-CVQKD follows the same trend from asymptotic limit to finite size limit GM-CVQKD. No positive finite size limit SKRs are achieved in LEO GM-CVQKD in bad atmospheric conditions. This is evidence of the inability of CVQKD to operate in bad atmospheric conditions due to significant signal attenuation. 

\begin{figure}[!htb]
    \centering
    \begin{subfigure}[b]{0.49\textwidth}
         \centering
         \includegraphics[width=\textwidth]{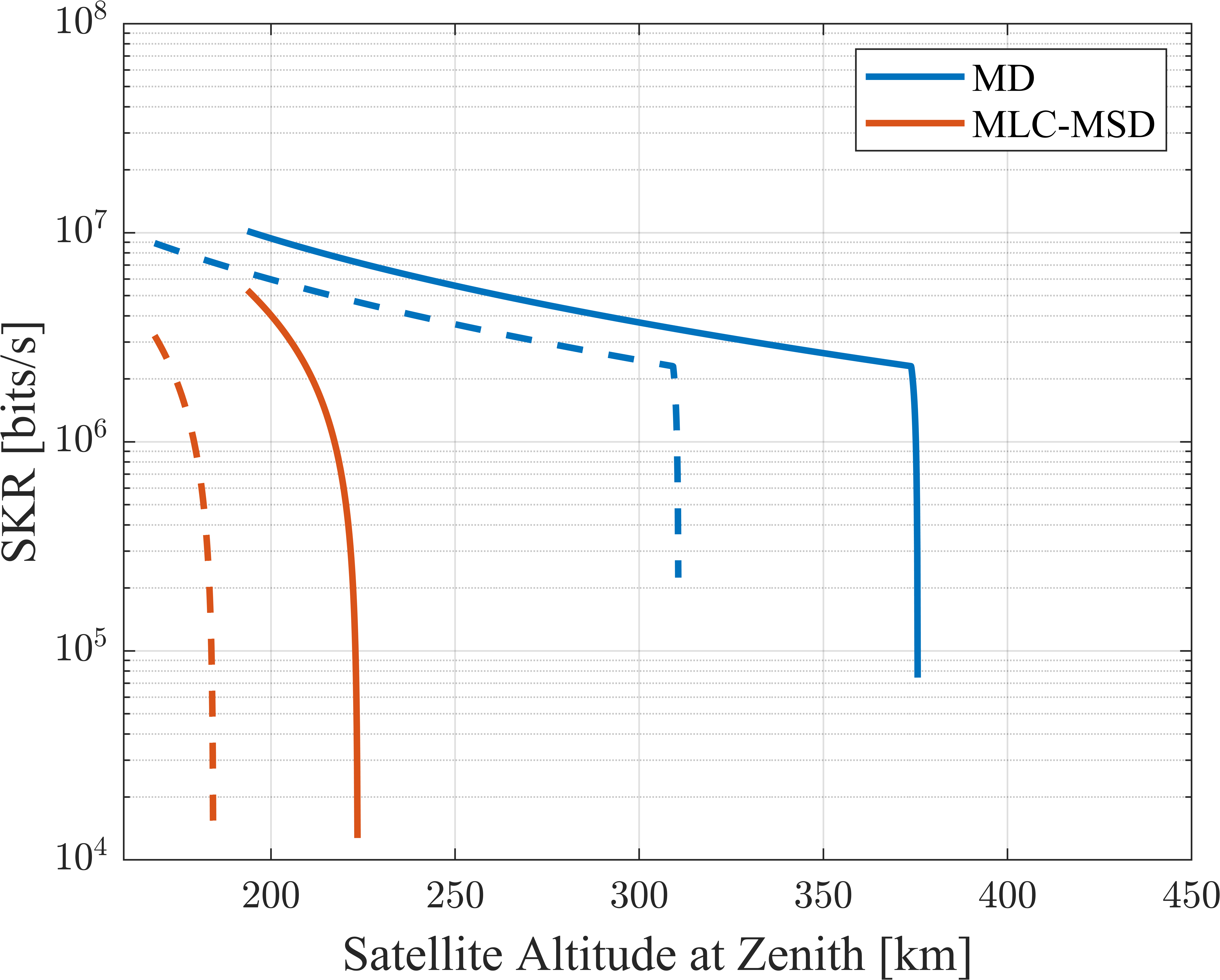}
         \caption{}
         \label{fig:F_Good_D1}
     \end{subfigure}
     \hfill
    \begin{subfigure}[b]{0.48\textwidth}
         \centering
         \includegraphics[width=\textwidth]{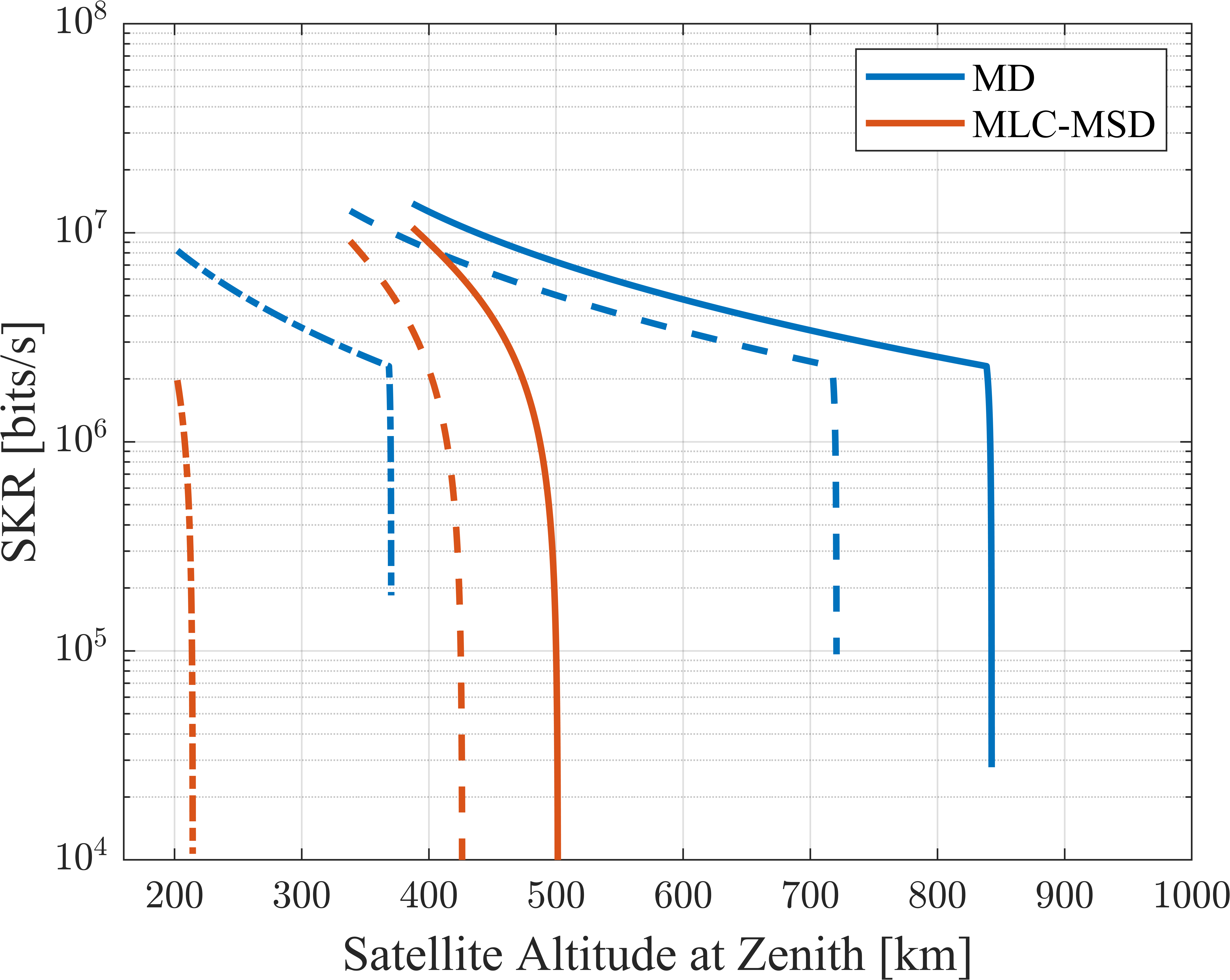}
         \caption{}
         \label{fig:F_Good_D2}
     \end{subfigure}
     \hfill
    \vspace{-2mm} 
    \caption{Finite size limit SKRs as a function of satellite altitude for a (a) 1m and (b) 2 m receiver aperture diameter GM-CVQKD downlink using MD (blue) and MLC-MSD (red) reconciliation using homodyne detection. The solid lines indicate $\theta =$ 90\degree, dashed lines indicate $\theta =$ 60\degree, dash-dotted lines indicate $\theta =$ 30\degree.}
    \label{fig:F_SKR_graph}
    \vspace{-8mm}
\end{figure}

Figure \ref{fig:F_SKR_graph} shows the finite SKRs in good atmospheric conditions for MD and MLC-MSD reconciliation. For a 1 m receiver aperture diameter, positive SKRs are restricted to a satellite orbit altitude below approximately 375~km and with larger elevation angles. Active propulsion is required to maintain this altitude, adding to the required complexities of a CVQKD LEO satellite. However, this situation can be ameliorated by increasing the receiver aperture diameter. In this case, the receiver aperture diameter was increased from 1 m to 2 m, increasing the satellite orbit altitude for which there is a positive SKR to a maximum of approximately 850~km and with lower elevation angles (Figure \ref{fig:F_Good_D2}). The results show that MD reconciliation outperforms MLC-MSD reconciliation by producing larger positive SKRs for larger link distances and lower elevation angles.

\begin{figure}[htp]
    \centering
    \includegraphics[width=0.49\textwidth]{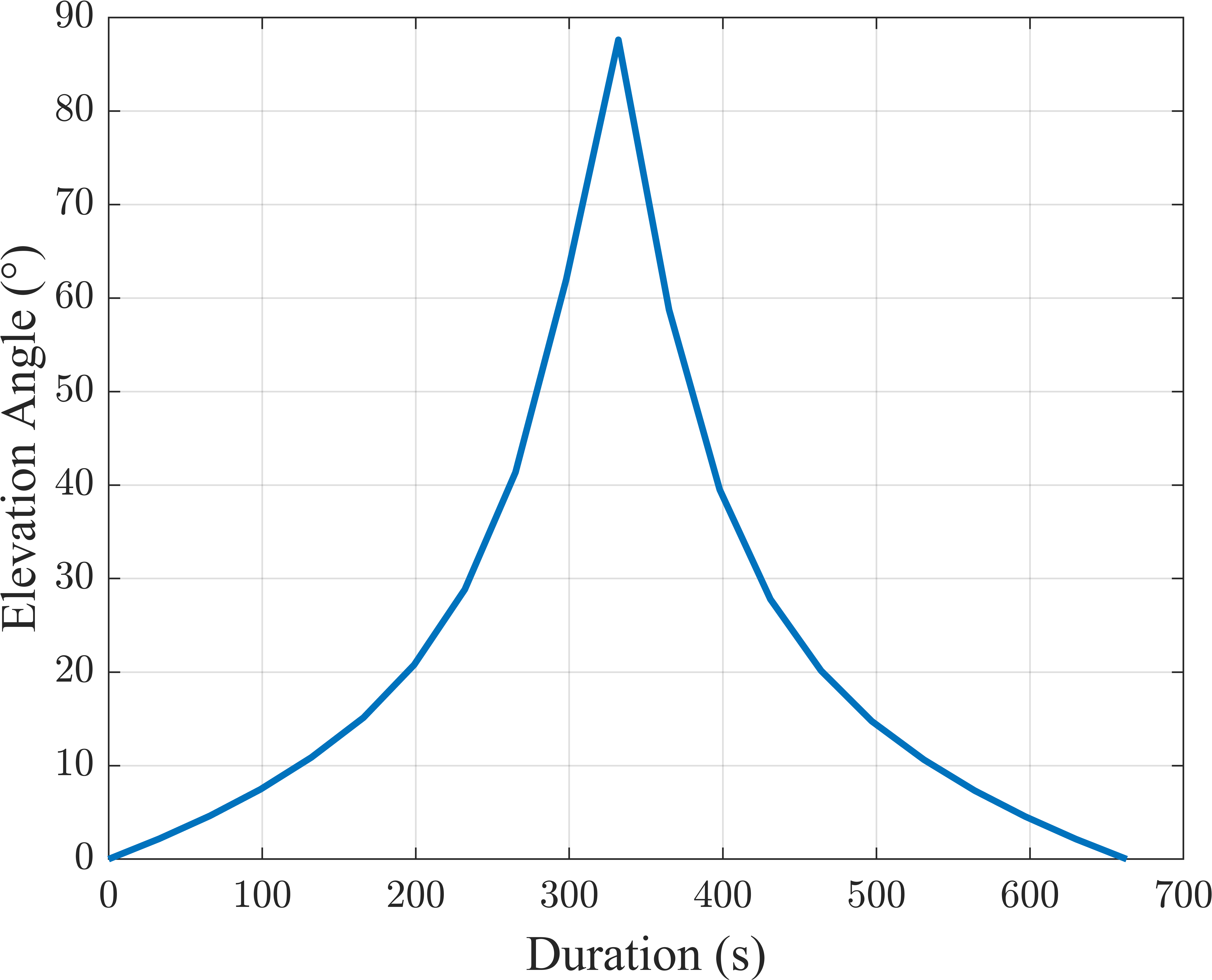}
    \caption{Elevation for an ISS pass over Mt. John Observatory on 9th August 2022. The pass had a maximum elevation angle of 87.6\degree.}
    \label{fig:ISS_Pass}
    \vspace{-5mm}
\end{figure}

The finite size limit SKR calculation can also be presented as a function of elevation angle for a satellite at a certain orbit altitude. For this purpose, we assessed the orbit of the International Space Station (ISS), which has an average orbit altitude at zenith of 417.5~km, and studied a pass, shown in Figure \ref{fig:ISS_Pass}, which occurred on 9th August 2022, over the University of Canterbury's Mt. John Observatory in New Zealand (Latitude = -43.9853\degree, Longitude = 170.4641\degree, Altitude = 1.029~km). The pass had a duration of 663~s with a maximum elevation angle of 87.6\degree{}. The calculations were made using the same link parameter values as in Table \ref{Parameters} with a receiver diameter aperture, $D_r$, of 2~m and OGS altitude, $L_{\mathrm{geo}}$, of 1.029~km. A practical limitation to be noted is that a LEO satellite pass with large elevation angles may go through the keyhole of an observatory. In this situation, the satellite cannot be tracked.

\begin{figure}[htp]
    \centering
    \includegraphics[width=0.49\textwidth]{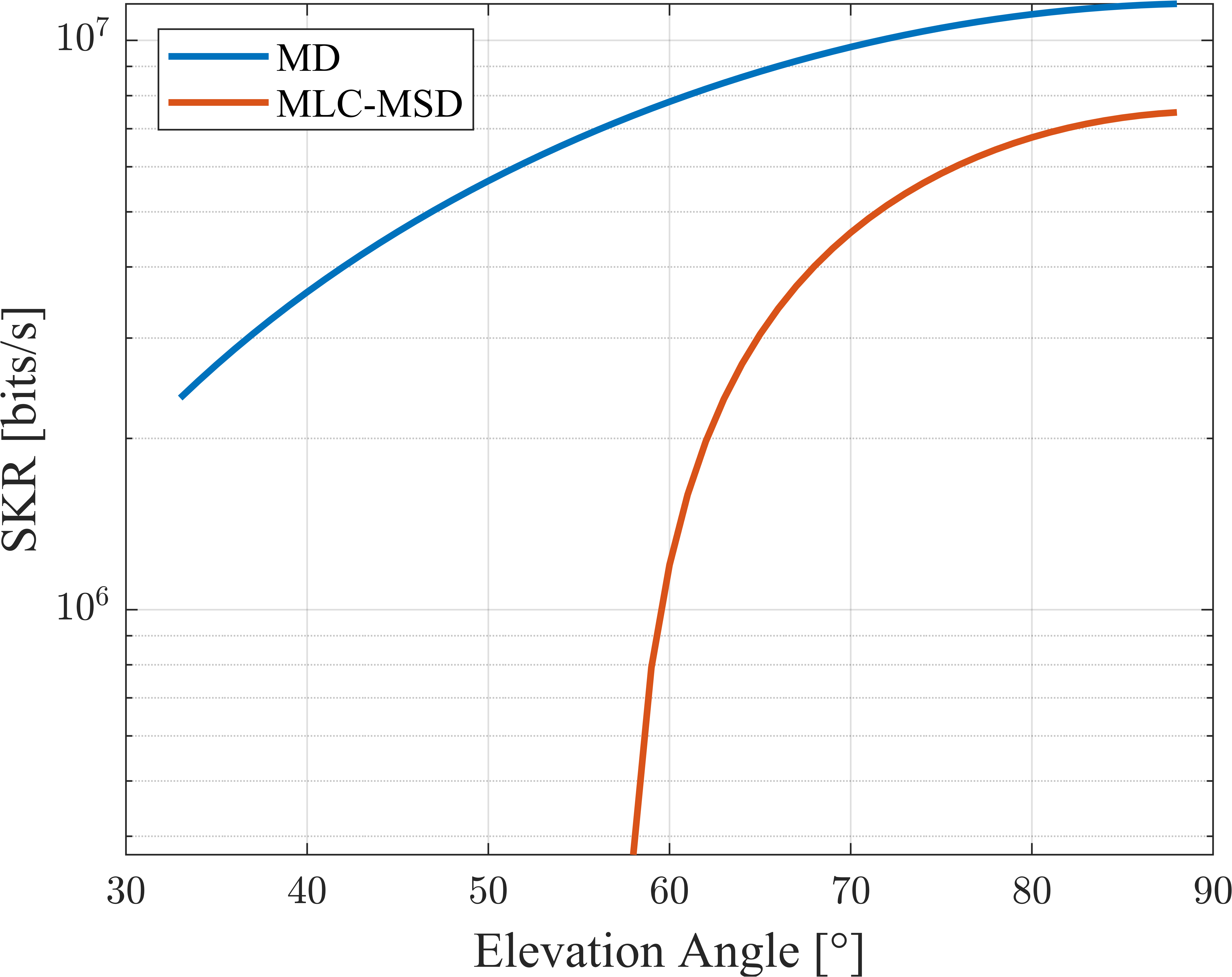}
    \caption{SKR for ISS pass over Mt. John Observatory with maximum elevation angle 87.6\degree. $D_r = $ 2~m. Homodyne detection. $L_{\mathrm{geo}} = $ 1.029~km.}
    \label{fig:ISS_Pass_SKR}
    \vspace{-5mm}
\end{figure}

The SKR was calculated assuming good atmospheric conditions for the ISS pass (Figure \ref{fig:ISS_Pass_SKR}). The results show that finite size limit GM-CVQKD with MD reconciliation outperforms MLC-MSD reconciliation by producing larger positive SKRs and at lower elevation angles.

Multiplying the calculated SKR and the time that the ISS is at a certain elevation angle produces the total number of bits of the secret key. The elevation angle has been discretised with a resolution of 1\degree{} in Figure~\ref{fig:ISS_Pass} to determine the temporal position of the ISS. The secret key has:
\begin{itemize}
    \item 1.235~Gbit for GM-CVQKD with MD reconciliation,
    \item 385~Mbit for GM-CVQKD with MLC-MSD reconciliation.
\end{itemize}

\subsection{SKR Trends from Parameter Variation Analysis}
A sensitivity analysis of the parameters leads to the following observed trends:
\begin{itemize}
    \item Increasing the laser repetition rate increases the finite size limit SKR.
    \item Decreasing the value for allowable information for Eve in the finite size limit ($\delta n_{\mathrm{privacy}}$) increases the SKR for a given link distance. This implies that there is greater accuracy in the estimated amount of information Eve has intercepted (Holevo information).
    \item Increasing the transmitter and receiver aperture diameter as well as the transmitter and receiver optics efficiencies increases the maximum link distance and therefore satellite orbit altitude. This increase results from the decrease in losses during state preparation and measurement.
    \item Decreasing the pointing loss increases the maximum link distance and therefore satellite orbit altitude. This requires a more accurate APT system as well as less beam wandering between Alice and Bob.
    \item Increasing the OGS elevation increases the maximum link distance and therefore satellite orbit altitude for which a positive SKR is possible. This results from the decrease in attenuation from a decrease in link distance and effective atmosphere thickness as Bob is effectively closer to Alice.
    \item Operating in good atmospheric conditions where there is greater visibility and a smaller refractive index structure parameter increases the maximum link distance and therefore satellite orbit altitude. This results from the decrease in attenuation due to less atmospheric scattering and turbulence.
    \item Protocols with a larger number of coherent states lead to larger SKRs.
\end{itemize}

\section{Conclusions}
In this work, a satellite-to-ground CVQKD channel link model for GM-CVQKD and DM-CVQKD in the asymptotic limit has been developed. The results show that GM-CVQKD outperforms both the $M$-PSK and $M$-QAM DM-CVQKD models in the asymptotic limit. However, the $M$-QAM protocol outperforms the $M$-PSK protocol and is capable of producing positive SKRs in both LEO and low MEO. In the finite size limit, MD reconciliation outperforms MLC-MSD reconciliation by producing larger 
positive SKRs in GM-CVQKD for larger link distances and lower elevation angles. The physical implementations of both GM-CVQKD and DM-CVQKD as well as the processes for MD and MLC-MSD reconciliation have not been considered. These should be investigated further for future satellite-to-ground and long distance CVQKD operation.

\bibliographystyle{IEEEtran}
\addcontentsline{toc}{section}{\refname}\bibliography{MSayat_SatelliteToGroundCVQKD}

\begin{IEEEbiographynophoto}{Mikhael Sayat} received the B.E. (Hons.) degree in Mechatronics Engineering and B.Sc. degree in Physics from the University of Auckland, New Zealand in 2021. He was the recipient of a 2022 New Zealand Space Scholarship to intern at NASA JPL. He is a doctoral candidate at the University of Auckland and is currently at the Department of Quantum Science and Technology, Australian National University. His research interests include quantum key distribution, quantum communications, quantum networks, and space-based quantum technologies.
\vspace{-25mm}
\end{IEEEbiographynophoto}


\begin{IEEEbiographynophoto}{Biveen Shajilal} received the M.Sc. degree in Photonics from the Cochin University of Science and Technology, Kerala, India. He is a doctoral candidate at the Department of Quantum Science and Technology, Australian National University. He is part of the ARC Centre of Excellence for Quantum Computation and Communication Technology. His research interests include quantum states of light, quantum communications, quantum entanglement and quantum information.
\vspace{-25mm}
\end{IEEEbiographynophoto}


\begin{IEEEbiographynophoto}{Sebastian P. Kish} was awarded a Ph.D. in Physics from The University of Queensland, Australia, in 2019. Sebastian is currently a postdoctoral fellow at The Australian National University and previously held a research associate position at UNSW. His research interests include quantum communications, quantum information and quantum entanglement.
\vspace{-25mm}
\end{IEEEbiographynophoto}


\begin{IEEEbiographynophoto}{Syed M. Assad} received the B.Sc. degree with a double major in physics and computational science from the National University of Singapore, and the joint Ph.D. degree in realization of harmonic entanglement between a light beam and its second harmonic, and theoretical proofs of security for quantum key distribution protocols from the National University of Singapore and The Australian National University in 2011. He was with the National University of Singapore as a Teaching Assistant and the Centre for Quantum Technologies as a Research Assistant. He has been with the Quantum Optics Group, The Australian National University, since 2011, where he is currently leading the Secure Communications Team. He is a member of the Centre of Excellence for Quantum Computation and Communication Technology.

\end{IEEEbiographynophoto}


\begin{IEEEbiographynophoto}{Thomas Symul} graduated from the Ecole Nationale Superieure des Telecommunications of Paris in 1998, and received is PhD degree in Physics from the University of Paris VI in 2001. He is currently the quantum research and development lead at QuintessenceLabs and one of its scientific co-founders. He has published over 50 papers during his career, including in prestigious journals such as Nature and Physical Review Letters, and has been awarded the Eureka Prize for Scientific Research in 2006 as well as the AIP Alan Walsh medal for service to the industry in 2014.
\vspace{-25mm}
\end{IEEEbiographynophoto}


\begin{IEEEbiographynophoto}{Ping Koy Lam} received a BSc degree with a double major in mathematics and physics from The University of Auckland, an MSc degree in theoretical physics and a PhD degree in experimental physics from The Australian National University (ANU). He was a Process Engineer with Sony (audio electronics) and Hewlett-Packard (semiconductor LED) for three years prior to his post-graduate studies with ANU. He was a recipient of the Australian Institute of Physics Bragg Medal and the Australian National University Crawford Prize for his PhD dissertation in 1999. He was awarded the British Council Eureka Prize for inspiring science (2003) and the University of New South Wales Eureka Prize for innovative research (2006). He is currently the chief quantum scientist at the Agency for Science, Technology, And Research (A*STAR) in Singapore. He has authored around 260 articles on his research in quantum physics.
\end{IEEEbiographynophoto}


\begin{IEEEbiographynophoto}{Nicholas Rattenbury} was awarded a Ph.D. in Physics from The University of Auckland, New Zealand, in 2004. Nicholas leads research in free space optical communications, astrodynamics and time-domain astrophysics.
\end{IEEEbiographynophoto}


\begin{IEEEbiographynophoto}{John Cater} was awarded a B.E. with Honours in Mechanical Engineering from the University of Auckland, New Zealand in 1997, and a Ph.D. from Monash University in Australia in 2003. John leads research in aerospace engineering; his interests include electric propulsion for small satellites and modelling of thermal protection systems for reentry of spacecraft.
\end{IEEEbiographynophoto}

\end{document}